\documentclass[journal]{IEEEtran}

\usepackage{url}
\usepackage{xurl}
\usepackage{cite}
\usepackage{amsmath,amssymb,amsfonts}
\usepackage{algorithmic}
\usepackage{graphicx}
\usepackage{textcomp}
\usepackage{xcolor}
\usepackage{hyperref}
\usepackage{stfloats}
\usepackage{booktabs}

\DeclareMathOperator{\relu}{ReLU}
\DeclareMathOperator{\softmax}{softmax}
\DeclareMathOperator{\clip}{clip}
\DeclareMathOperator{\hnrop}{HNR}
\newcommand{\un}[1]{\,\mathrm{#1}}
\newcommand{\Nm}{\ensuremath{\mathrm{N{\cdot}m}}}
\newcommand{\tr}{^{\mathsf{T}}}

\newcommand{\fs}{f_{\mathrm{s}}}

\newcommand{\preemph}{\alpha}
\newcommand{\frameidx}{t}
\newcommand{\nframes}{T}

\newcommand{\freq}{f}
\newcommand{\win}{\ell}
\newcommand{\spec}{X}
\newcommand{\specw}{\tilde{X}}
\newcommand{\gauss}{G}

\newcommand{\sigw}{\sigma_{\mathrm{w}}}
\newcommand{\sigc}{\sigma_{\mathrm{c}}}

\newcommand{\speed}{\omega}
\newcommand{\speedhat}{\hat{\omega}}
\newcommand{\accel}{\dot{\omega}}
\newcommand{\accelhat}{\dot{\hat{\omega}}}
\newcommand{\frot}{f_{\mathrm{rot}}}
\newcommand{\frothat}{\hat{f}_{\mathrm{rot}}}
\newcommand{\order}{k}
\newcommand{\orderset}{\mathcal{K}}
\newcommand{\ncyl}{n}
\newcommand{\ncylhat}{\hat{n}}
\newcommand{\ncylset}{\mathcal{N}}
\newcommand{\ncylsetsix}{\mathcal{N}_{6}}
\newcommand{\cand}{r}
\newcommand{\candset}{\mathcal{R}}
\newcommand{\binw}{\delta}

\newcommand{\combkern}{\psi}
\newcommand{\comb}{E}
\newcommand{\combmat}{\mathbf{E}}
\newcommand{\combref}{C}
\newcommand{\combrefmat}{\mathbf{C}}
\newcommand{\evid}{A}
\newcommand{\evidmat}{\mathbf{A}}
\newcommand{\evidvec}{\mathbf{a}}
\newcommand{\gain}{g}
\newcommand{\gainvec}{\mathbf{g}}
\newcommand{\detfeat}{\mathbf{d}}

\newcommand{\readwa}{\mathbf{W}_{\mathrm{a}}}
\newcommand{\readwd}{\mathbf{W}_{\mathrm{d}}}
\newcommand{\readbias}{\mathbf{b}}
\newcommand{\readwh}{\mathbf{W}_{\mathrm{h}}}
\newcommand{\readbiash}{\mathbf{b}_{\mathrm{h}}}
\newcommand{\hid}{\mathbf{h}}
\newcommand{\readout}{\mathbf{q}}
\newcommand{\invtemp}{\beta}
\newcommand{\score}{s}
\newcommand{\post}{P}
\newcommand{\loss}{J}

\newcommand{\countcrit}{Q}
\newcommand{\vel}{v}
\newcommand{\stiffa}{\kappa_{\mathrm{a}}}
\newcommand{\stiffv}{\kappa_{\mathrm{v}}}
\newcommand{\level}{L}

\newcommand{\torque}{\tau}
\newcommand{\torquehat}{\hat{\tau}}
\newcommand{\torquewot}{\tau_{\mathrm{WOT}}}
\newcommand{\torquedrag}{\tau_{\mathrm{drag}}}
\newcommand{\torqueload}{\tau_{\mathrm{L}}}
\newcommand{\inertia}{\Theta}
\newcommand{\load}{\lambda}
\newcommand{\loadhat}{\hat{\lambda}}
\newcommand{\loadcoast}{\lambda_{0}}
\newcommand{\loadstate}{u}
\newcommand{\loadstatehat}{\hat{u}}

\newcommand{\feat}{\mathbf{z}}
\newcommand{\bandprof}{B}
\newcommand{\hnr}{\hnrop}
\newcommand{\lambins}{D}
\newcommand{\lambinc}{\lambda}
\newcommand{\siglam}{\sigma_{\lambda}}

\newcommand{\relerr}{\epsilon}

\def\BibTeX{{\rm B\kern-.05em{\sc i\kern-.025em b}\kern-.08em
  T\kern-.1667em\lower.7ex\hbox{E}\kern-.125emX}}

\begin{document}

\title{SonoDyno: A Sonic Dynamometer for Engine Speed and Load Estimation from Audio}

\author{Robin Doerfler and Beat Rossmy
\thanks{This work has been submitted to the IEEE for possible publication. Copyright may be transferred without notice, after which this version may no longer be accessible.}
\thanks{The authors are with Audio Research \& Development, Impulse Audio Lab GmbH,
Munich, Germany (e-mail: robin.doerfler@impulse-audio-lab.com;
beat.rossmy@impulse-audio-lab.com).}
}

\maketitle
\IEEEpeerreviewmaketitle

\begin{abstract}
Controllable engine sound synthesis, Active Sound Design and engine order analysis all build on the relationship between the sound of an engine and its momentary operating state: its speed and its load. Recordings of engine sound are abundant in archives, sample packs and online video, yet they rarely carry time-aligned annotations of that state, which limits their applicability to controllable engine sound generation. We present SonoDyno, which estimates the engine speed, the cylinder count and an engine-normalized load intensity from a microphone signal alone, with no tachometer, no bus access and no prior knowledge of the engine configuration. A convolutional scorer of 124\,302 parameters reads the energy at the engine orders, the harmonics of half the rotation frequency, in each frame at three time-frequency resolutions to jointly score cylinder count and speed, and a decoder imposes the rotational dynamics of the engine on the speed trajectory. A second head estimates the load intensity, the position of the torque between the engine's drag floor at 0 and its full-load ceiling at 1, from the order levels at the decoded speed. On recordings of engines it has never heard, with microphone position, engine type and cylinder count unknown, SonoDyno places 93 to 97\,\% of the frames within 2\,\% of the true speed, identifies the cylinder count with 80\,\% accuracy and confuses the octave on fewer than 2\,\% of the frames. On an unseen engine of a known type, the load intensity has a mean absolute error of 0.12, and the sign of the torque, driving or braking, is right on 91\,\% of the frames. General pitch trackers fall behind even when the harmonic order of each engine is supplied: on our validation recordings and at a tolerance of 5\,\%, the best of them places 62\,\% of the frames within the tolerance, against 96\,\% for SonoDyno.
\end{abstract}

\begin{IEEEkeywords}
RPM estimation, torque estimation, engine load, cylinder count, order analysis, pitch estimation, engine sound synthesis.
\end{IEEEkeywords}

\section{Introduction}
\label{sec:intro}

The sound of a combustion engine follows its momentary operating state: its speed (how fast it turns) and its load (how hard it works to produce the demanded torque).
Order analysis of engine noise and vibration needs the speed as its reference~\cite{wangVehicleNoiseVibration2010, luTacholessSpeedEstimation2019, peetersReviewComparisonTacholess2019}, and engine sound synthesis relies on both for analysis and control: from driving simulators~\cite{heitbrinkDesignDrivingSimulation2007} and real-time engine simulation~\cite{yaghiEngineSimulator2022} to the sound inside the vehicle itself, where Active Sound Design synthesizes engine harmonics and noise components to shape the perceived character of a combustion vehicle or to give the driver of an electric vehicle acoustic feedback on the drivetrain state~\cite{boddenPrinciplesActiveSound2016, kimSystematicApproachEngine2017, doerflerGradientBasedLearning2026, caoEngineOrderSound2020, zhangResearchActiveSound2023, kimSmartphoneActiveSoundDesign2024}.
Supervised generative approaches in particular depend on these annotations for control and conditioning~\cite{lundbergDataDrivenProceduralAudio2020, lobatoMotor2Synth2025, choiDDSPNeuralVehicle2026, doerflerPhysicsInformedNeural2026}.
However, measuring the operating state requires specialized instrumentation: a tachometer, a dynamometer or access to the often proprietary messages on the vehicle's Controller Area Network (CAN) bus~\cite{choiDDSPNeuralVehicle2026}. Annotated recordings therefore stay scarce.
While procedural generation can expand a few of them into large corpora, its seed recordings must still carry time-aligned control traces themselves~\cite{doerflerAnalysisDrivenProceduralGeneration2026}.
In contrast, unannotated engine recordings are abundant in archives, sample libraries, field recordings and online sources, and reading their operating state from the audio would make them available to all of these methods.

For speed estimation, the established methods fall into two families:
Tacholess speed estimation derives the rotational speed of a machine from its vibration or sound~\cite{luTacholessSpeedEstimation2019, peetersReviewComparisonTacholess2019, huaARC2026}, and pitch estimation tracks the fundamental frequency of speech and music, using classical signal processing~\cite{decheveigneYIN2002, camachoSWIPE2008, mauchPYIN2014} or machine learning~\cite{kimCREPE2018, riouPESTO2023}.
Neither addresses the load.

Pitch estimation is a natural candidate, since engine sound shares the source-filter structure of the voice: a train of pulses shaped by a resonant system. But the pulses differ:
where the voice produces one glottal pulse per period, a four-stroke engine produces one exhaust pulse from each of its cylinders in a cycle of two crankshaft revolutions.
The sound therefore repeats every two revolutions, and its harmonics, the \emph{engine orders}, lie at multiples of half the rotation frequency.
The pulses arrive at the firing frequency, the cylinder count times the cycle rate, and identical cylinders would produce only the order at that frequency, the firing order, and its multiples.
The orders in between, which reveal the rotation itself, arise only from differences between the cylinders and stay weaker.
How strongly each order sounds depends on the shape of the pulses and on the exhaust system they pass through (Section~\ref{subsec:signal}).

Engine recordings depart from what both families assume in three ways.
First, the speed cannot be inferred from the position of the harmonics on the frequency axis alone.
The strongest line lies at the firing frequency, and since each cylinder fires once every two revolutions, that frequency is half the cylinder count times the rotation frequency: an engine with half the cylinders at twice the speed fires at the same frequency.
A four-cylinder engine at $6000\un{rpm}$ and an eight-cylinder engine at $3000\un{rpm}$ both fire at $200\un{Hz}$, and their dominant lines coincide (Section~\ref{subsec:signal}).
Reading the speed from the harmonic comb therefore requires the cylinder count, which has to be estimated together with the speed.
Pitch estimators return one periodicity per frame, the firing frequency or one of its octaves, and leave the count open.
Tacholess methods take the order from the gear ratios and other known design data of the machine~\cite{peetersReviewComparisonTacholess2019}, and Motor2Synth, which adapts a pitch estimator to engine audio, applies an order chosen for each clip~\cite{lobatoMotor2Synth2025}.
None of them reads the count from the sound, and a recording of an unknown engine offers no other source for it.
Second, the analysis faces a conflict of resolution.
Near idle, at $600\un{rpm}$, adjacent orders lie $5\un{Hz}$ apart and can only be separated with an analysis window of $200\un{ms}$ or longer, yet when the engine revs freely or a gear changes, the speed changes by more than $1500\un{rpm/s}$, and within such a window the firing order of an eight-cylinder engine sweeps $20\un{Hz}$.
The estimator therefore needs fine resolution in both frequency and time.
Third, the harmonic structure changes drastically even within one engine.
As the speed rises, the orders sweep through the resonances of the exhaust system, which act on the pulse train as a comb filter~\cite{doerflerPhysicsInformedNeural2026}, and the hotter gas under load moves these resonances~\cite{tekinExperimentalInvestigationEffects2022}.
Between full load and overrun, the excitation itself changes, from strong combustion pulses rich in high orders to an engine that only pumps air with the fuel cut off (Section~\ref{subsec:signal}).
The exhaust system, the engine bay and the cabin each filter the pulses differently, so the microphone position reshapes the spectrum, and a turbocharger, tire and wind noise or the rollers of a dynamometer add significant confounding noise (Section~\ref{subsec:signal}).
Recordings thus capture an ensemble of machinery and transfer paths, and the estimator must extract harmonic cues that indicate the speed regardless of these confounding factors.

In synthesis applications, the engine torque is available as an input, an absolute quantity in \Nm{}.
It already reflects what the vehicle asks of the engine, since mass, gear and gradient determine the torque needed for the requested motion, but its scale is specific to the engine: the same torque is a light load on a large engine and a heavy one on a small engine.
What the sound carries is the \emph{effort} of the engine, its torque relative to its own capacity, on a continuum from engine braking in overrun, with the fuel cut off, through coasting to full load.
The absolute torque therefore cannot be recovered from a recording (Section~\ref{subsec:signal}).
Lacking it, engine sound synthesis methods trained on recordings with differentiable digital signal processing (DDSP) resort to loudness as a proxy for the effort~\cite{engelDDSPDifferentiableDigital2020, lobatoMotor2Synth2025}.
Loudness, however, carries neither the sign nor the scale of the effort.
Acceleration and overrun at the same loudness have torques of opposite sign and thus sound different, and the same loudness on two different engines does not imply the same effort.
We therefore make the effort itself the target.
We define the \emph{load intensity} $\load \in [0, 1]$, which places the current torque between two curves of the engine over speed: its \emph{drag floor}, the negative torque in overrun with the fuel cut off, at 0, and its \emph{full-load ceiling} at 1.
With these two curves, torque and $\load$ convert into each other at every speed: an application converts its measured torque to $\load$ to condition a synthesis model, and a predicted $\load$ rescales to \Nm{}.

SonoDyno reads each frame through three analysis windows, from long to short, so that both the close orders at idle and their fast sweeps in a gear change are resolved.
From each window we obtain its \emph{order energy}, the spectral energy at the positions of all engine orders for each candidate speed of a logarithmic grid.
A learned scorer refines and fuses this energy per order and scores every pair of cylinder count and speed, so that count and speed are estimated together.
The count enters only through a \emph{firing-order gain}, weights that set which orders to expect strong for that count, so that one model serves the eleven cylinder counts of our data, from 1 to 18.
A decoder then keeps the speed on a physically plausible path, since its acceleration follows the net torque, which changes slowly, and it reads the cylinder count of the recording from the best path.
A separate head estimates the load intensity from the levels of the orders at the decoded speed.

Our contributions are:
\begin{enumerate}
  \item joint estimation of cylinder count and engine speed from a microphone signal alone, in which the count acts only through a firing-order gain over the engine orders, so that one model serves eleven counts, including counts with a single training engine (Sections~\ref{subsec:scoring} and~\ref{subsec:results_rpm});
  \item a front end of order energy at three resolutions, refined within each window and fused across windows per order without mixing the orders, beside an order-pattern detector, which reads the pattern across all orders without a count and carries the scorer through fast speed changes (Sections~\ref{subsec:frontend} to~\ref{subsec:scoring} and~\ref{subsec:design_choices});
  \item a rotational-dynamics decoder, a Viterbi search over the speed and its relative rate of change that penalizes changes of that rate (Section~\ref{subsec:decoder});
  \item an engine-normalized load intensity as the target and a head that estimates it from the order levels at the decoded speed (Sections~\ref{subsec:recording} and~\ref{subsec:loadhead}).
\end{enumerate}

We evaluate on real engines held out whole, on held-out sessions of a training engine and on simulated engines never seen, and evaluate the test split once.
On real engines it has never heard, SonoDyno places 93 to 97\,\% of the frames within 2\,\% of the true speed and identifies the cylinder count with 80\,\% accuracy.
On the real validation recordings, the best general pitch tracker places 62\,\% of the frames within 5\,\% of the true speed even when the harmonic order is supplied, against 96\,\% for SonoDyno (Section~\ref{sec:results}).
Its scorer, with 124\,302 learned parameters, is no larger than the smallest learned tracker.
On an unseen engine of a known type, the load intensity (0 to 1) has a mean absolute error of 0.12.
Section~\ref{subsec:design_choices} measures the design choices one by one.
A conference paper presents the order energy, the decoder and the load-intensity head for an acoustics audience~\cite{doerflerRPMTorqueAAC2026}; this paper adds the learned scorer, the cylinder-count read and the measurement of their design choices.

\section{Background and Related Work}
\label{sec:background}

\subsection{Engine Sound as a Signal}
\label{subsec:signal}

\textbf{Engine orders:} In a four-stroke engine, one engine cycle spans two revolutions of the crankshaft and holds one exhaust pressure pulse from each of the $\ncyl$ cylinders~\cite{jonesModellingExhaustNoise1984, doerflerPhysicsInformedNeural2026}. The pulse sequence therefore repeats every two revolutions, and its repetition gives rise to harmonic components, the engine orders, which lie at multiples of half the crankshaft rotation frequency $\frot = \speed/60$, with $\speed$ the engine speed in rpm: order $\order$ lies at $\order\frot$, with $\order \in \{\tfrac12, 1, \tfrac32, \dots\}$~\cite{wangVehicleNoiseVibration2010, puderSpeechEnhancementHandsFree2003, cevherVehicleSpeedEstimation2009}. Identical, evenly spaced firings produce only the firing order $\ncyl/2$ and its multiples, at the firing frequency $\tfrac{\ncyl}{2}\frot = \ncyl\speed/120$, $\ncyl$ times the fundamental $\frot/2$ of the signal~\cite{wangVehicleNoiseVibration2010}. Differences from cylinder to cylinder, in pulse pressure and shape, firing interval or exhaust path, repeat once per cycle and fill the orders in between, mostly weaker than the firing orders~\cite{wangVehicleNoiseVibration2010, leeHalfOrderModulation2019}. The speed thus places the orders, the cylinder count decides which of them dominate, and the shape of the pulses and the exhaust system set their amplitudes~\cite{jonesModellingExhaustNoise1984, doerflerPhysicsInformedNeural2026}.

\textbf{Count and speed:} The firing frequency $\ncyl\speed/120$ is set by the product of count and speed alone, so all pairs with the same product share it together with all of its harmonics. Only the weaker orders in between tell such pairs apart: they lie $\frot/2 = \speed/120$ apart, half as far for the engine with twice the cylinders at half the speed. A tracker that follows the strongest line therefore recovers the firing frequency, and inferring the correct speed from it requires $\ncyl$. Under one assumed count, the confusion appears as an octave error, and across counts it moves the speed by the ratio of the two counts.

\textbf{Resolution:} Low speeds and fast speed changes set opposite demands on the analysis window: separating adjacent orders needs a window of at least $120/\speed$ seconds, the inverse of their spacing, which is about $200\un{ms}$ at $600\un{rpm}$, the bottom of the range we consider; the Blackman windows of our front end (Section~\ref{subsec:frontend}), whose main lobe reaches its first zeros at $\pm 3/\win$ for a window of length $\win$, need three times that, about $600\un{ms}$. A speed change at the rate $\accel$, in rpm per second, moves order $\order$ by $\order\accel/60$ hertz per second, so when the engine revs freely or a gear changes, the orders sweep across several spacings within such a window. Since the sweep rate grows with the order, in the same frame a low order may need the long window and a high order the short one, and the need flips within a run-up.
The front end therefore analyzes each frame at several window lengths (Section~\ref{subsec:frontend}). Within one window a speed change smears every order, yet all orders move by the same factor, so their common motion is visible across the orders and in no single one.

\textbf{Load:} The load changes the pulses themselves, and with them the order spectrum, even at a constant speed~\cite{doerflerPhysicsInformedNeural2026}, and under full load the sound grows most in its upper range~\cite{tekinExperimentalInvestigationEffects2022}.
The order spectrum also becomes denser: on six of our seven real engines at 1500 to $3000\un{rpm}$, high load leaves 1.3 to 1.8 times as many orders at least $6\un{dB}$ above the floor between them as overrun does.
The exhaust resonances follow the speed of sound in the gas, which rises with its temperature~\cite{tekinExperimentalInvestigationEffects2022}, so the hotter gas under load shifts them up in frequency.
In overrun, with the fuel cut off, there is no combustion: the engine only pumps air~\cite{doerflerPhysicsInformedNeural2026}, and all orders sink towards the noise floor, so fewer of them clear it.
Between the two, the balance between the orders, and between harmonic and noise energy, follows the load~\cite{doerflerGradientBasedLearning2026, zhangResearchActiveSound2023}.

\textbf{Torque and effort:} Referred to the crankshaft, the drivetrain of the vehicle obeys
\begin{equation}
  \inertia\,\tfrac{2\pi}{60}\,\accel = \torque - \torqueload,
  \label{eq:drivetrain}
\end{equation}
with $\torque$ the engine torque, $\inertia$ the effective inertia of the engine and of the vehicle mass reflected through the engaged gear, $\torqueload$ the load torque referred to the crankshaft, and $2\pi/60$ the conversion of $\accel$ from rpm per second to radians per second squared.
An estimator obtains $\speed$ from the positions of the orders and $\accel$ from their course over time (Sections~\ref{subsec:frontend} and~\ref{subsec:decoder}). The sound does not reveal the inertia $\inertia$ or the load torque $\torqueload$, which change with the mass of the vehicle, the engaged gear and the gradient of the road. The acceleration thus shows only the sign of the net torque $\torque - \torqueload$, whether the engine torque lies above or below the load torque, and at a steady speed the two are equal. Since the net torque changes slowly, so does $\accel$, which the decoder exploits (Section~\ref{subsec:decoder}). The effort behind a torque is also relative to the capacity of the engine. A four-cylinder and an eight-cylinder engine of the corpus peak at $304\un{\Nm{}}$ and $803\un{\Nm{}}$ (Fig.~\ref{fig:lambda}), so $300\un{\Nm{}}$ is full load for the first and below 40\,\% of the peak of the second. The sound therefore carries only the effort, the position of the torque in the engine's own range, from its drag floor $\torquedrag(\speed)$ to its full-load ceiling $\torquewot(\speed)$, the torque at wide-open throttle, which Section~\ref{subsec:recording} defines as the load intensity $\load$.

\textbf{Transfer path and other sources:} The speed and load cues reach the microphone through a transfer path and among other sources. At the exhaust outlet the pulses have passed the resonators and silencers of the exhaust system, and in the cabin also the vehicle body~\cite{wangVehicleNoiseVibration2010}. Outside the vehicle, the body can suppress the firing frequency in some directions~\cite{cevherVehicleSpeedEstimation2009}. The order weighting and dynamic range of the same engine state therefore change with the microphone position. Other sources add to the engine: tire and wind noise are broadband~\cite{puderSpeechEnhancementHandsFree2003, hadleyNoiseReductionDriverPitCrew2006}, a turbocharger adds a tone at its own rotation speed~\cite{gagliardiTurbochargerRotational2019}, and a dynamometer adds the noise of its rollers. The microphone position and the broadband noise change the spectral envelope and the absolute level, while the frequencies of the orders stay fixed. Whitening each spectrum removes the envelope and makes the harmonic cues of one engine state comparable across microphone positions (Section~\ref{subsec:frontend}).

\subsection{Tacholess Speed Estimation and Order Analysis}
\label{subsec:tacholess}

Order analysis needs the instantaneous speed~\cite{saavedraAccurateAssessmentComputed2006}, and where no tachometer can be fitted it is estimated from the measured vibration or sound itself~\cite{luTacholessSpeedEstimation2019, schmidtTacholessOrderTracking2018}. The established estimators track a ridge in the time-frequency plane, in one variant predicting it under constant angular acceleration from a harmonic and a starting frequency the user supplies~\cite{schmidtTacholessOrderTracking2018}, or refine a coarse ridge estimate by demodulation~\cite{urbanekTwoStepProcedure2013}. Others demodulate
the phase of one dominant harmonic while the speed fluctuates little~\cite{bonnardotUseAccelerationSignal2005}. Vold-Kalman filters then extract individual orders, given the speed and the order number~\cite{voldHighResolutionOrder1993, gadeCharacteristicsVoldKalman1999}. A comparison of these estimators on vibration data found that each depends on inputs tuned to the data set, and that fast speed changes and operating changes that alter the harmonic structure remain open~\cite{peetersReviewComparisonTacholess2019}. Speed has also been read from airborne sound, each time for one known machine. One study takes the firing frequency from a microphone and an accelerometer on the bonnet of one car, given its cylinder count~\cite{shanAnalysisVibrationAcoustic2020}. Another tracks the tone of a turbocharger with a frequency-locked loop~\cite{gagliardiTurbochargerRotational2019}. A third sorts the speed of one car into classes with a learned classifier that did not transfer to another car~\cite{fosalauEngineRotationalSpeed2022}.
On the vibration side, a learned regressor maps gearbox vibration onto speed~\cite{beeringEstimationRotationalSpeeds2022}. A recent tracker, ARC, maps the outputs of YIN, the cepstrum and a harmonic comb onto a shared uniform speed grid with a fixed order~\cite{huaARC2026}. It keeps their octave alternatives as separate modes and relies on a motion-prior tracker to leave a wrong octave. All of these methods assume a known machine and are evaluated on the machine they were fitted to: the order is chosen by hand or fixed, the sensor is usually a contact accelerometer.

\subsection{Pitch and Periodicity Estimation}
\label{subsec:pitch}

Classical estimators find one periodicity per frame from a candidate function: the cepstrum~\cite{nollCepstrumPitchDetermination1967}, the harmonic product spectrum~\cite{schroederPeriodHistogram1968}, subharmonic summation on a logarithmic frequency axis~\cite{hermesMeasurementPitchSubharmonic1988}, the cumulative-mean-normalized difference function of YIN~\cite{decheveigneYIN2002} and the sawtooth template of SWIPE~\cite{camachoSWIPE2008}. Subharmonic summation is the classical counterpart of the order energy that Section~\ref{subsec:frontend} computes per order on a logarithmic speed grid. pYIN turns frame-wise candidates into a trajectory with a hidden Markov model decoded by the Viterbi algorithm~\cite{mauchPYIN2014, viterbi1967}. Its transitions penalize changes of pitch, a first-order prior. The decoder of Section~\ref{subsec:decoder} penalizes changes of the rate of speed change. Supervised networks classify pitch on a logarithmic output axis~\cite{kimCREPE2018}, or build harmonic structure into the input: the harmonic constant-Q transform stacks harmonics as channels on a shared logarithmic axis~\cite{bittnerDeepSalience2017}. The order-pattern detector of Section~\ref{subsec:scoring} takes the orders as channels in the same way. Self-supervised estimators learn pitch from transposition: SPICE from the known shift between two pitch-shifted views~\cite{gfellerSPICE2020}, PESTO from transposition equivariance~\cite{riouPESTO2023, riouPESTORealTime2025}. Both determine pitch up to a constant offset, the anchor, which PESTO fixes with a shift calibrated on synthetic sounds of known pitch. A neural pitch estimator trained on speech or on music fails on the other where its training data do not cover the pitches it meets, and serves both once they do~\cite{morrisonCrossDomainNeuralPitch2023}.
Engine sound falls outside this family on two points. First, the strongest line, which an estimator reads as the period, lies at the firing frequency, $\ncyl$ times the fundamental of the signal. Second, the anchor that a self-supervised estimator needs is the order of that line, which depends on the unknown cylinder count.

\subsection{Engine Audio: Synthesis Demand and Prior Estimators}
\label{subsec:engineaudio}

Engine synthesizers are conditioned on speed, and most of them on load or a proxy for it: sample-based and granular playback indexed by speed~\cite{heitbrinkDesignDrivingSimulation2007, jaglaSamplebasedEngineNoise2012, chenSynthesisingSoundCar2021, liRealTimeAutomotiveEngine2024}, procedural and physically informed models~\cite{farnellDesigningSound2010, baldanPhysicallyInformedCar2015}, DDSP~\cite{engelDDSPDifferentiableDigital2020, lundbergDataDrivenProceduralAudio2020} and pulse-train, gain-curve and procedural models~\cite{doerflerPhysicsInformedNeural2026, doerflerGradientBasedLearning2026, doerflerAnalysisDrivenProceduralGeneration2026}. Every data-driven one needs control trajectories for its training material, annotated or estimated. Choi et al.\ condition a DDSP vehicle synthesizer on CAN signals of one vehicle~\cite{choiDDSPNeuralVehicle2026}, and synthesis quality generally rises as driving signals are added. For their four-cylinder engine, a fundamental at the rotation frequency outperforms one at the firing frequency, which the authors attribute to the components between the firing harmonics.

The cylinder count has been classified from phone recordings as one vehicle attribute among several~\cite{terwilligerImprovingMisfireFault2022}, and a drive-by study names it the most elusive engine parameter~\cite{cevherVehicleSpeedEstimation2009}. A thesis classifies the count first and then ranks firing-frequency candidates under it, from synchronized sound and vibration of engines at steady speed~\cite{kimPhysicsInformedTwoStage2026}.
A classifier over the count, however, needs training engines for every class, and some counts are rare (Section~\ref{subsec:coverage}).

Motor2Synth adapts a pitch estimator to engine audio~\cite{lobatoMotor2Synth2025}: it retrains PESTO self-supervised on sample packs of nine vehicles, chooses an order for each sample, and smooths the result with a hidden Markov model and a moving average. Load enters its synthesis as loudness. Its reported accuracy is measured on samples of the vehicles its estimator was trained on, against ground truth from a commercial speed-extraction tool and with the order chosen per sample, so it cannot be compared with ours. We therefore reproduce its strategy on our data, retraining PESTO with the published self-supervised recipe on our training split, and benchmark it beside stock PESTO, CREPE, pYIN and YIN. Every tracker receives an oracle order chosen on the labels, which favors the trackers (Section~\ref{sec:results}).

Outside sound synthesis, vibro-acoustic condition monitoring reads the combustion of an engine from its vibration and sound~\cite{delvecchioVibroacousticCondition2018}: cylinder pressure has been reconstructed from acoustic-emission sensors on the engine block~\cite{jafariInCylinderPressure2021}, and combustion parameters have been regressed from a knock sensor on the cylinder head~\cite{kefalasEstimationCombustionParameters2022}. Both rely on sensors mounted on laboratory research engines.
The tacholess, pitch and engine-audio literatures thus stop at the same question, which line belongs to which order, and the load has been read only through contact sensors or as loudness.

\section{Method}
\label{sec:method}

SonoDyno is an offline annotator. It takes the microphone channels of a complete recording session and returns, for each microphone and every frame $\frameidx$ of a $64\un{ms}$ grid, the engine speed $\speedhat_\frameidx$ and the load intensity $\loadhat_\frameidx$, the latter also averaged over the microphones of the session. For the session, it returns one cylinder count $\ncylhat$, read from all microphones together.
The count is chosen among eleven hypotheses $\ncyl \in \ncylset = \{1, 2, 3, 4, 5, 6, 8, 10, 12, 16, 18\}$.

\begin{figure*}[t]
\centering
\includegraphics[width=\textwidth]{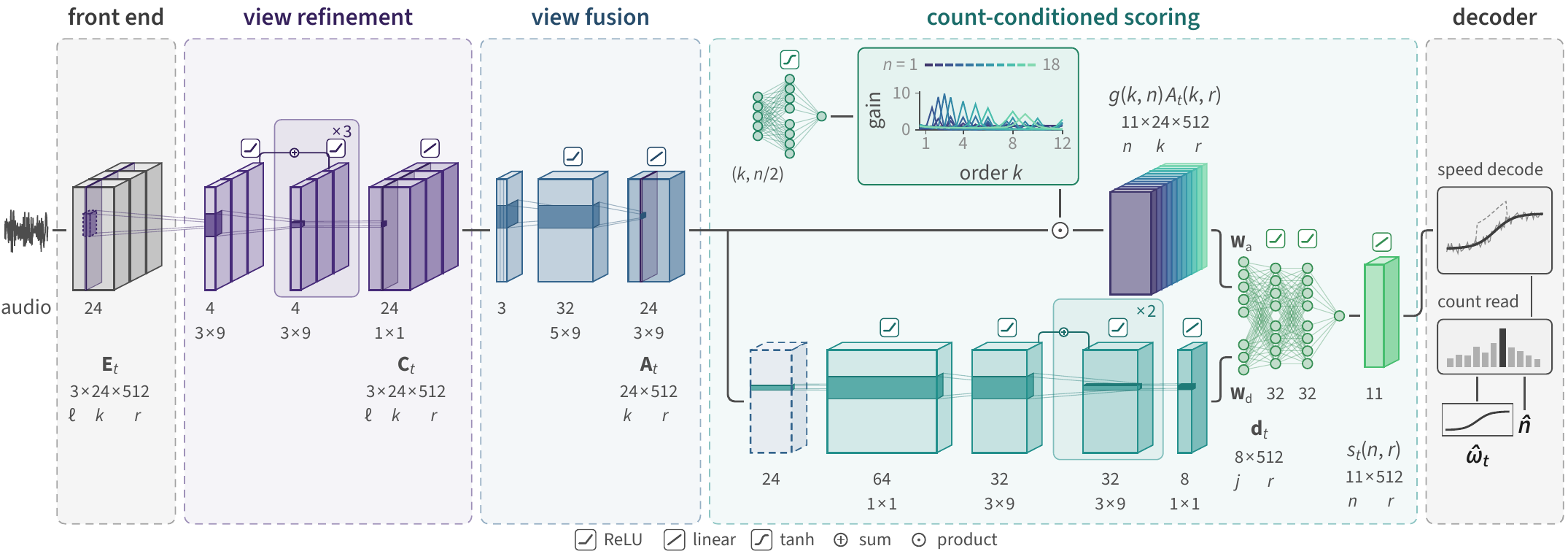}
\input{figures/fig_architecture.caption}
\label{fig:architecture}
\end{figure*}

A fixed front end turns each frame into order energy $\combmat_\frameidx$: for every analysis window, or view, every order and every candidate speed, the energy the frame holds where that order would lie if the engine turned at that speed. Three learned stages follow. Stage 1, \emph{view refinement}, cleans the energy of each order within each window. Stage 2, \emph{view fusion}, merges the windows of each order into one map, the \emph{fused evidence}. Stage 3 scores every pair of count and candidate speed: a firing-order gain weights the orders for each count, an order-pattern detector reads all orders without a count, and a reader turns both into a score. Together the three stages form the scorer (Fig.~\ref{fig:architecture}), and one objective trains them together. A decoder without learned parameters imposes the rotational dynamics of the engine on the speed of each microphone and reads the count. A separate head estimates the load intensity.

Two invariants hold along the chain. Stages 1 and 2 never mix two orders, so the evidence reaches stage 3 as one map per order. No parameter belongs to a single count: one shared map computes the gain of every count, and the detector and the reader serve all counts alike, so a count acts on the score only through the weights its gain puts on the orders. Both follow from the signal model: the count decides which orders are strong, so it enters only where the orders are weighted.

The \emph{order comb}, the order set, the candidate grid and the physics features of each pair of order and firing order are fixed, while the cleaning of a ridge, the choice of window, the map from those features to the gain, the pattern across the orders and the reader are learned. The stages that do not depend on the count run once per frame, and only the reader works on every count, in one pass over all of them with the count as one more axis of its input.

The scorer holds 124\,302 learned parameters, 112\,552 of them (91\,\%) in the order-pattern detector.

\subsection{Front End}
\label{subsec:frontend}

The audio is resampled to $\fs = 8\un{kHz}$ and pre-emphasized with $\preemph = 0.97$, and this band holds order 12, the highest order the comb reads, which reaches $1800\un{Hz}$ at $9000\un{rpm}$.
Each frame is then analyzed through three Blackman windows of length $\win \in \{128, 512, 2048\}\un{ms}$, centered on the same frame. The 128 and $512\un{ms}$ windows are zero-padded to four times their length, so the 512 and $2048\un{ms}$ windows share bins of $0.488\un{Hz}$ and the $128\un{ms}$ window has bins of $1.95\un{Hz}$.
Each magnitude spectrum $\spec_\frameidx(\freq, \win)$ is normalized to its maximum.

\begin{figure}[t]
\centering
\includegraphics[width=\columnwidth]{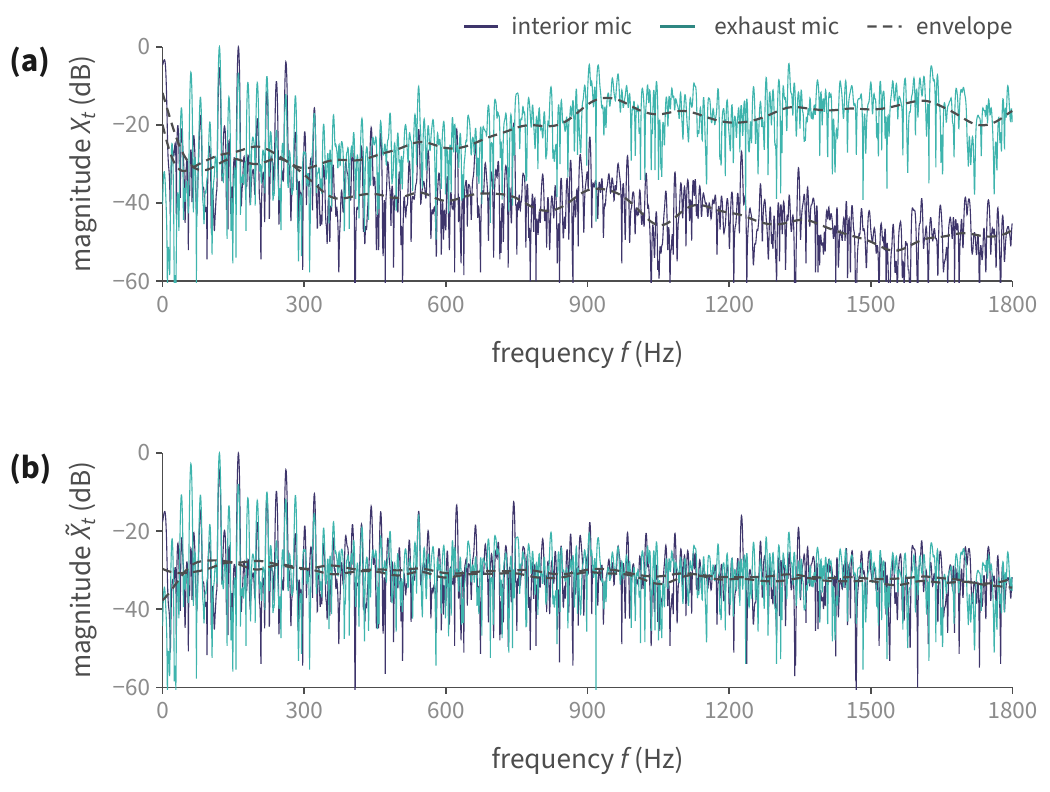}
\input{figures/fig_whitening.caption}
\label{fig:whitening}
\end{figure}

The microphone position shapes the spectral envelope and leaves the order lines in place. Whitening therefore removes the envelope of each frame, the log-magnitude smoothed by a Gaussian $\gauss$ of width $\sigw = 29\un{Hz}$ on every window, and keeps the order lines (Fig.~\ref{fig:whitening}),
\begin{multline}
  \log \specw_\frameidx(\freq, \win) = \log \spec_\frameidx(\freq, \win) \\
  - \sum_{\freq'} \gauss(\freq - \freq', \sigw)\, \log \spec_\frameidx(\freq', \win),
  \label{eq:whitening}
\end{multline}
followed by a second max normalization.

The speed axis on which the comb reads the whitened spectrum is a grid $\candset$ of 512 candidate speeds $\cand$, spaced logarithmically from 600 to $9000\un{rpm}$. One step is $\binw = 0.00765$ octave, $0.531\,\%$ of the speed, and an octave spans 131 candidates at every speed, so an octave error, or a ramp at a given relative rate, has the same shape anywhere on the grid.

\begin{figure}[t]
\centering
\includegraphics[width=\columnwidth]{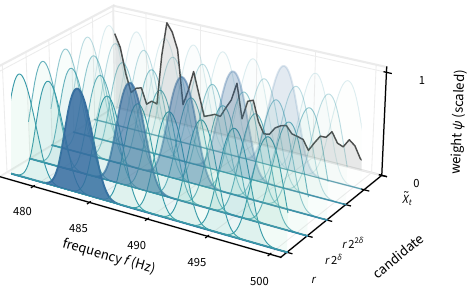}
\input{figures/fig_tiling.caption}
\label{fig:tiling}
\end{figure}

On this grid, for order $\order$ and candidate $\cand$, the comb kernel $\combkern(\freq, \win, \order, \cand)$ is a box of unit mass over the frequencies at which order $\order$ lies for the speeds of the grid cell of $\cand$, its edges blurred by a Gaussian of $\sigc = 1.5$ bins. The kernels of all orders and candidates form the order comb. The order energy sums the whitened spectrum under each kernel,
\begin{equation}
  \comb_\frameidx(\win, \order, \cand) = \sum_{\freq} \combkern(\freq, \win, \order, \cand)\, \specw_\frameidx(\freq, \win),
  \label{eq:orderenergy}
\end{equation}
and the kernels of neighboring candidates tile the frequency axis (Fig.~\ref{fig:tiling}), so the harmonic of every speed inside a cell falls under the kernel of that cell, and the candidate axis has no gaps. The comb runs on all three windows, over the 24 orders $\orderset = \{\tfrac12, 1, \tfrac32, \dots, 12\}$, and $\combmat_\frameidx$ holds $3 \times 24 \times 512$ values per frame.

\begin{figure}[t]
\centering
\includegraphics[width=\columnwidth]{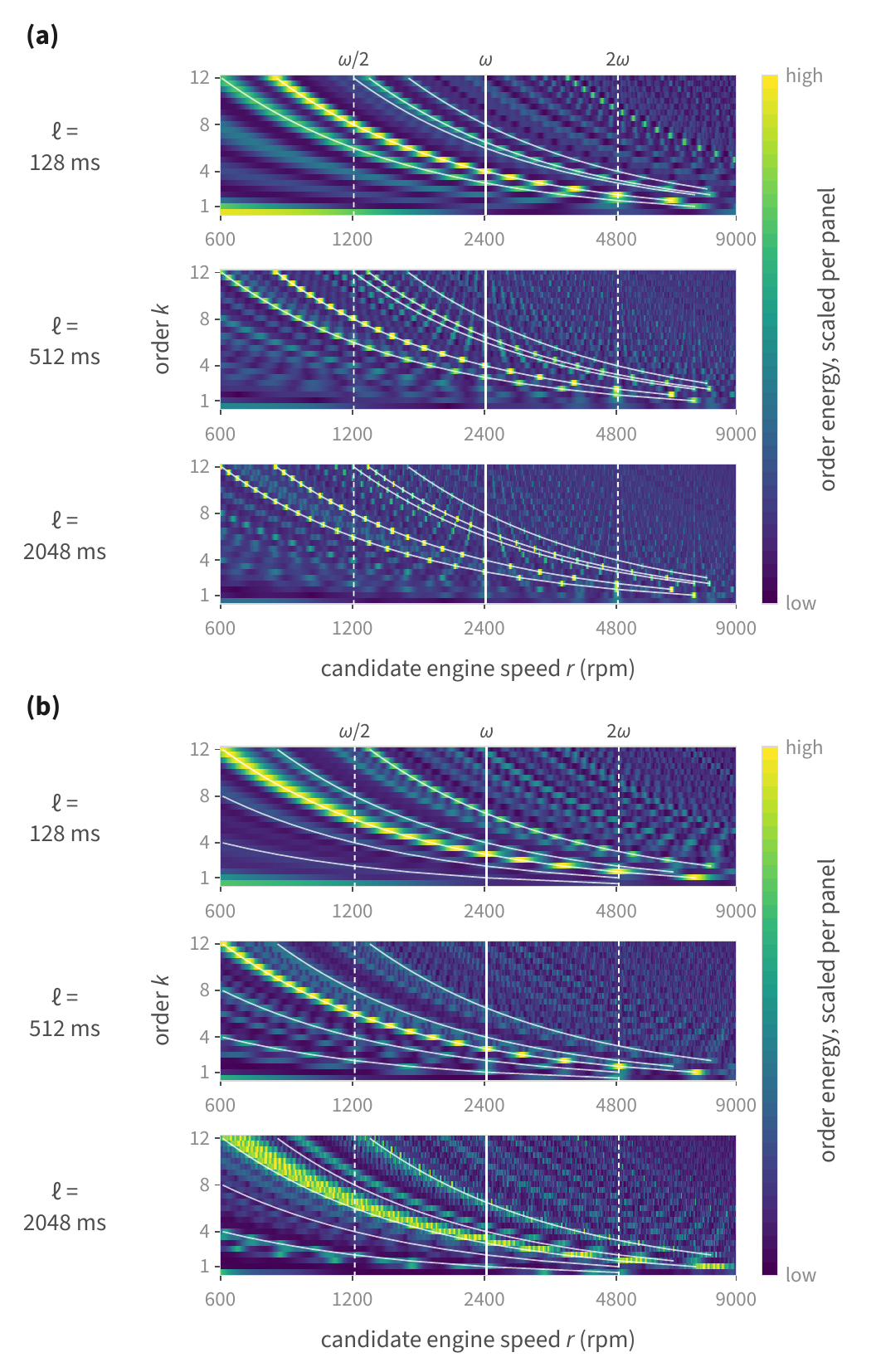}
\input{figures/fig_views.caption}
\label{fig:views}
\end{figure}

The three windows give the order energy three resolutions, and each serves a different regime: the $2048\un{ms}$ window separates the orders at low, steady speed, the $128\un{ms}$ window follows a ramp, and the $512\un{ms}$ window covers the cases between (Fig.~\ref{fig:views}). The learned stages start from these three views of every order.

\subsection{Stage 1: View Refinement}
\label{subsec:refine}

Within one window, the order energy of one order over frames and candidates forms its \emph{order map}. The order map shows a ridge along the speed, blurred by the leakage of the window, by the frequency bins, which at low orders span several candidates, and by noise. View refinement cleans this ridge. Since each window draws the ridge with its own geometry (Fig.~\ref{fig:views}), view refinement gives each window $\win$ its own stack of convolutions over (frame, candidate), which maps each order map to the \emph{refined order map} $\combref_\frameidx(\win, \order, \cand)$. On the logarithmic grid, a change of speed moves every order by the same number of candidates, so the 24 orders of a window share its stack.
The stack holds an input convolution into four channels with a rectified linear unit (ReLU), three residual blocks that each add the ReLU of a convolution to their input, and a linear $1 \times 1$ convolution back to one map. The other kernels are $3 \times 9$ over frames and candidates.
The stack sees $\pm 4$ frames and $\pm 16$ candidates.
The refined order maps $\combrefmat_\frameidx$ have the shape of $\combmat_\frameidx$.

\subsection{Stage 2: View Fusion}
\label{subsec:fusion}

Since the window that resolves an order changes with the order and over time, view fusion chooses the window locally and per order. The three refined order maps of one order, $\combref_\frameidx(\win, \order, \cdot)$ for the three windows, enter one small convolution stack as input channels, each window with its own weights. A convolution over $5 \times 9$ frames and candidates maps them into 32 hidden channels with ReLU. A linear convolution over $3 \times 9$ then gives one signed map, the fused evidence $\evid_\frameidx(\order, \cand)$. Over all orders and candidates it forms $\evidmat_\frameidx$.
The stack sees $\pm 3$ frames and $\pm 8$ candidates and is shared by the 24 orders. The choice of window acts as a switch: the short window during a ramp, the long one at steady speed. A fixed linear blend of the windows cannot express a switch. The hidden layer can, since each hidden channel responds to one combination of the windows, such as a clean ridge in the short window with a smear in the long one.

\subsection{Stage 3: Count-Conditioned Scoring}
\label{subsec:scoring}

\begin{figure}[t]
\centering
\includegraphics[width=\columnwidth]{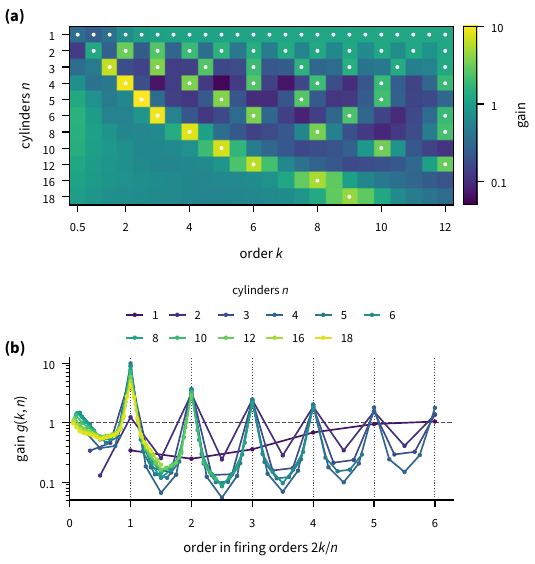}
\input{figures/fig_gain.caption}
\label{fig:gain}
\end{figure}

Stage 3 has three parts: the firing-order gain, the order-pattern detector and the reader. A count $\ncyl$ predicts which orders are strong: the firing order $\ncyl/2$ and its multiples. The firing-order gain turns this prediction into one weight per order. Five features describe each pair of order $\order$ and firing order $\ncyl/2$: the ratio $2\order/\ncyl$, its distance to the nearest integer, a soft indicator of that distance being small, and the base-2 logarithms of the two orders. One multilayer perceptron (MLP) with 16 tanh units maps the features to a logit, and a softmax over the orders gives
\begin{equation}
  \gain(\order, \ncyl) = 24\,\softmax_{\order}\big(\phi(\order, \ncyl/2)\big),
  \label{eq:gain}
\end{equation}
with $\phi$ the perceptron, so the weights of each count are positive and average 1 over the orders. The mean of 1 leaves each count a shape over the orders and no scale of its own. One map serves every count, so a count with one or two training engines (Section~\ref{subsec:coverage}) is weighted by the same map as a count with more than a hundred.
The result is an $11 \times 24$ table, constant over frames and candidates (Fig.~\ref{fig:gain}), and $\gainvec(\ncyl)$ collects the weights of count $\ncyl$.

Beside the gain, the order-pattern detector reads the pattern across all orders without a count. During a ramp all orders move together, and only a stage that sees all orders at once can use that. The 24 orders of $\evidmat_\frameidx$ enter as channels: a pointwise mix into 64 channels with ReLU, a convolution over $3 \times 9$ into 32 channels with ReLU, two residual blocks, and a $1 \times 1$ projection to the detector features $\detfeat_\frameidx(\cand)$, eight per frame and candidate. The detector sees $\pm 3$ frames and $\pm 12$ candidates of the fused evidence.

With $\evidvec_\frameidx(\cand)$ the column of $\evidmat_\frameidx$ at candidate $\cand$, one value per order, the reader turns the weighted evidence and the detector features into one score per count and candidate through two hidden layers of 32 units. The first is
\begin{multline}
  \hid_\frameidx(\ncyl, \cand) = \relu\big(\readwa\,(\gainvec(\ncyl) \odot \evidvec_\frameidx(\cand)) \\
  + \readwd\,\detfeat_\frameidx(\cand) + \readbias\big),
  \label{eq:reader}
\end{multline}
with $\readwa \in \mathbb{R}^{32 \times 24}$, $\readwd \in \mathbb{R}^{32 \times 8}$ and $\readbias \in \mathbb{R}^{32}$, and the second gives the score
\begin{equation}
  \score_\frameidx(\ncyl, \cand) = \invtemp\,\readout\tr \relu\big(\readwh\,\hid_\frameidx(\ncyl, \cand) + \readbiash\big),
  \label{eq:score}
\end{equation}
with $\readwh \in \mathbb{R}^{32 \times 32}$, $\readbiash, \readout \in \mathbb{R}^{32}$ and a learned inverse temperature $\invtemp$. The reader is pointwise in frame and candidate. Since $\readwa$ is signed, the gain sets how much the evidence of an order matters for a count, and the reader sets whether it raises or lowers the score. In a weighted sum, each order adds to the score independently of the other orders, whereas a hidden layer can make the response to one order depend on another: order 2 strong with order 1 weak separates four cylinders from two.

Across its three stages, no parameter of the scorer is indexed by the candidate. Every learned stage is a convolution or a pointwise map along the candidate axis, and the gain is constant over the candidates. The scorer is therefore shift-equivariant along the grid: a shift of the order energy shifts the scores by the same number of candidates, apart from the ends of the grid, and it has no parameter through which to learn a prior over speed.

\subsection{Objective}
\label{subsec:objective}

A single softmax over the joint grid $\ncylset \times \candset$, the $11 \times 512$ cells of a frame, turns the scores into the posterior
\begin{equation}
  \post_\frameidx(\ncyl, \cand) = \frac{\exp \score_\frameidx(\ncyl, \cand)}
  {\sum_{\ncyl' \in \ncylset} \sum_{\cand' \in \candset} \exp \score_\frameidx(\ncyl', \cand')}.
  \label{eq:posterior}
\end{equation}
The objective $\loss$ is the cross-entropy of the labeled cell: the mean over the labeled frames of $-\log \post_\frameidx(\ncyl, \cand)$, with $\ncyl$ the cylinder count of the engine and $\cand$ the candidate nearest the labeled speed $\speed_\frameidx$. Frames outside 600 to $9000\un{rpm}$ and frames with the engine off carry no target.
Each labeled frame thereby lowers the probability of the true count at every wrong speed, the octave errors included, and the probability of every other count at every speed. The count is thus learned together with the speed, as one coordinate of a single score.

\subsection{Rotational-Dynamics Decoder and Count Read}
\label{subsec:decoder}

\begin{figure}[t]
\centering
\includegraphics[width=\columnwidth]{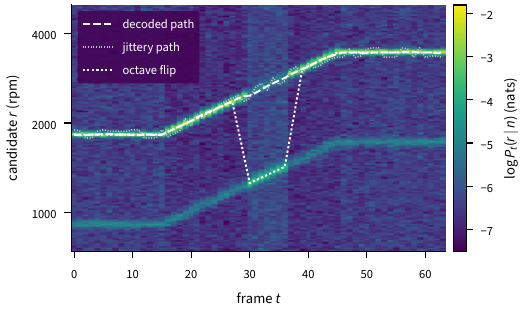}
\input{figures/fig_decoder.caption}
\label{fig:decoder}
\end{figure}

An octave spans 131 candidates, more than the scorer sees around any candidate, so continuity over a microphone's signal belongs to the decoder. The acceleration follows the net torque (Eq.~\eqref{eq:drivetrain}), which changes slowly against a $64\un{ms}$ frame, so a path that keeps its rate of change is physical, and jitter or a jump by an octave is not. The rotational-dynamics decoder therefore charges changes in the relative rate of speed change.
The decoder works on the per-count posterior $\post_\frameidx(\cand \mid \ncyl) = \post_\frameidx(\ncyl, \cand) / \sum_{\cand'} \post_\frameidx(\ncyl, \cand')$. A path $(\cand_1, \dots, \cand_\nframes)$ runs over the $\nframes$ frames of one microphone, and its \emph{path velocity} $\vel_\frameidx = (\log_2 \cand_\frameidx - \log_2 \cand_{\frameidx-1})/\binw$, in candidates per frame, is the relative rate of speed change. With weights $\stiffa$ on the change of path velocity and $\stiffv$ on the path velocity, the decoder finds
\begin{multline}
  \arg\max_{\cand_1, \dots, \cand_\nframes} \sum_{\frameidx=1}^{\nframes} \log \post_\frameidx(\cand_\frameidx \mid \ncyl)
  - \stiffa \sum_{\frameidx=3}^{\nframes} \big((\vel_\frameidx - \vel_{\frameidx-1})\,\binw\big)^2 \\
  - \stiffv \sum_{\frameidx=2}^{\nframes} (\vel_\frameidx\,\binw)^2
  \label{eq:decoder}
\end{multline}
by a Viterbi search on the state (candidate, path velocity).
The term $(\vel_\frameidx - \vel_{\frameidx-1})\,\binw$ is the second difference of $\log_2 \cand_\frameidx$ in octaves per frame squared, the change in that rate, so one $\stiffa$ holds at every speed of the grid.
The speed decode runs under the count that the \emph{count read} below takes from the session, $\ncyl = \ncylhat$, with $\stiffa = 512$, $\stiffv = 0$ and $\lvert \vel_\frameidx \rvert \le 48$: a constant path velocity within the bound costs nothing (Fig.~\ref{fig:decoder}).
The estimate $\speedhat_\frameidx$ is the mean of the candidates within $\pm 3$ candidates of the path, weighted by $\post_\frameidx(\cand \mid \ncylhat)$.

\begin{figure}[t]
\centering
\includegraphics[width=\columnwidth]{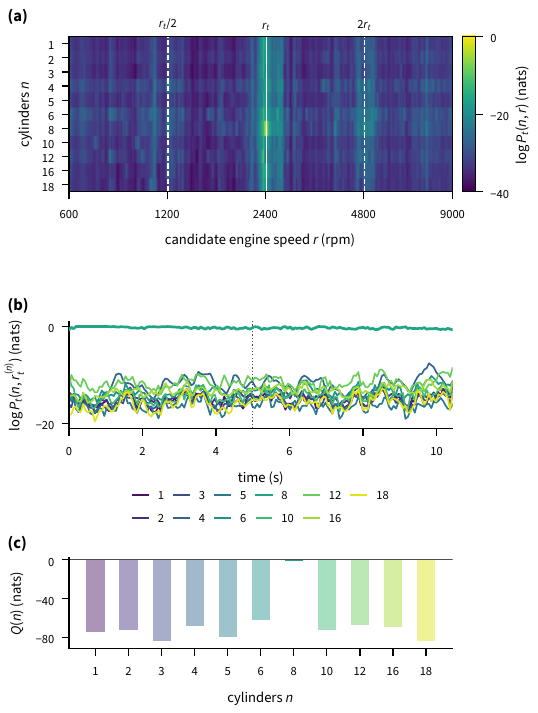}
\input{figures/fig_joint.caption}
\label{fig:joint}
\end{figure}

The count read, the session-level estimate $\ncylhat$, decodes one path $\cand_\frameidx^{(\ncyl)}$ for every count from its own posterior $\post_\frameidx(\cand \mid \ncyl)$, with the first-order setting of Eq.~\eqref{eq:decoder}: $\stiffa = 0$, $\stiffv = 32$ and no bound on the path velocity. It scores each path under the joint posterior,
\begin{multline}
  \countcrit(\ncyl) = \sum_{\text{microphones}} \frac{1}{\nframes} \sum_{\frameidx}
  \log \post_\frameidx\big(\ncyl, \cand_\frameidx^{(\ncyl)}\big) \\
  \text{and} \qquad \ncylhat = \arg\max_{\ncyl \in \ncylset} \countcrit(\ncyl),
  \label{eq:countread}
\end{multline}
the outer sum running over the microphones of the session and $\nframes$ being the number of frames of each (Fig.~\ref{fig:joint}). The joint posterior of a cell is its per-count posterior times the probability of count $\ncyl$ in that frame, so $\countcrit(\ncyl)$ weighs the evidence for the count as well as the fit of its path.

An engine-off gate sets $\speedhat_\frameidx = 0$ where the engine is off. The broadband level $\level_\frameidx$ of the raw audio, smoothed by a zero-phase exponential moving average over $300\un{ms}$, marks a frame as off when it lies both $12\un{dB}$ below the 90th percentile of that microphone's level and below $-44\un{dB}$ relative to full scale, each with $6\un{dB}$ of hysteresis.

\subsection{Load-Intensity Head}
\label{subsec:loadhead}

\begin{figure}[t]
\centering
\includegraphics[width=\columnwidth]{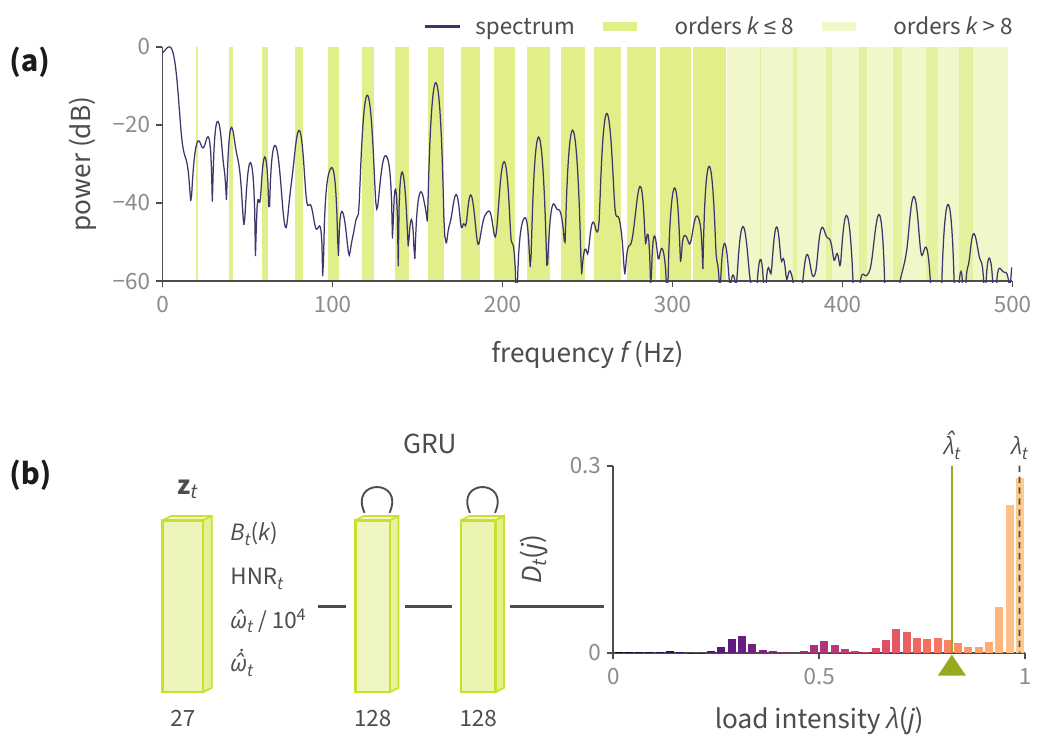}
\input{figures/fig_lambda_head.caption}
\label{fig:lambdahead}
\end{figure}

Whitening and max normalization discard the level and the envelope of each frame, and with them the two properties of the order spectrum that change with load (Section~\ref{subsec:signal}): how many orders stand above the floor between them, and how the energy divides between the orders and that floor. The load-intensity head therefore reads the raw power spectrum $\lvert \spec_\frameidx(\freq, 512) \rvert^2$ of the $512\un{ms}$ window, placed at the decoded speed $\speedhat_\frameidx$ of the microphone (Fig.~\ref{fig:lambdahead}).

Its input $\feat_\frameidx$ holds 27 values. The band profile $\bandprof_\frameidx(\order)$ gives, for each of the 24 orders, the power in dB within $\pm 3\,\%$ of $\order\,\frothat$, where $\frothat = \speedhat_\frameidx / 60$ is the decoded rotation frequency.
The harmonic-to-noise ratio $\hnr_\frameidx$ sets the energy in the bands of the orders up to 8 against the energy between those bands, in dB.
The last two values are the speed $\speedhat_\frameidx / 10^4$ and its rate of change $\accelhat_\frameidx$.
The head sees nothing of the scorer's internal state.

A causal gated recurrent unit (GRU) with two layers of 128 units reads the standardized sequence $\feat_1, \dots, \feat_\frameidx$ and returns at every frame a distribution $\lambins_\frameidx(j)$ over 40 equal bins of $\load$ with centers $\lambinc(j)$, and the probability $\loadstatehat_\frameidx$ that the torque is positive.
The estimate is the center of mass of that distribution,
\begin{equation}
  \loadhat_\frameidx = \sum_{j=1}^{40} \lambins_\frameidx(j)\, \lambinc(j).
  \label{eq:loadhat}
\end{equation}
The center of mass suits a distribution that is broad and often has one mode in overrun and one under load for the same spectrum: its peak, taken as the estimate, explains less of the variance of $\load$ on validation.
The head is trained with the cross-entropy of $\lambins_\frameidx$ against a Gaussian target of width $\siglam = 1$ bin centered on the labeled $\load_\frameidx$, plus 0.3 times the binary cross-entropy of $\loadstatehat_\frameidx$ against the load state $\loadstate_\frameidx$, which is 1 where the torque is positive and 0 in overrun.

Frames the gate closes read $\loadhat_\frameidx = 0$. Where the engine's drag floor and full-load ceiling are known, the estimate converts to torque as $\torquehat_\frameidx = \torquedrag(\speedhat_\frameidx) + \loadhat_\frameidx\, (\torquewot(\speedhat_\frameidx) - \torquedrag(\speedhat_\frameidx))$.
When a session has several microphones, $\loadhat_\frameidx$ and $\loadstatehat_\frameidx$ are averaged per frame over the microphones on which the engine runs, so that one load reading describes the session.

\section{Data and Supervision}
\label{sec:data}

\subsection{Corpus}
\label{subsec:corpus}

The corpus pairs engine audio with sample-aligned labels of engine speed and engine torque: 116.7\,h at 48\,kHz.
It combines three kinds of material, summarized in Table~\ref{tab:corpus}. Real recordings contribute 41.9\,h from 8 engines in 538 recording sessions, with cylinder counts 4, 6 and 8 and peak torques from 304 to 803\,\Nm{}.
Simulated material contributes 60.4\,h from 370 engines with 11 cylinder counts from 1 to 18. It is rendered by video-game engine simulations, which use either sample-based synthesis from recordings or physical modeling of the gas dynamics~\cite{yaghiEngineSimulator2022}. Most of these engines are rendered under the same 15 operating protocols,
 from constant load and run-ups to dynamic drive cycles.
The third kind, procedural resyntheses, adds 14.4\,h, a subset of a published procedural dataset~\cite{doerflerProceduralEngineSoundsDataset2025, doerflerAnalysisDrivenProceduralGeneration2026}. They resynthesize the sound of two of the real engines along speed and torque trajectories recorded on real vehicles, so their labels are recorded telemetry.

\begin{table}[t]
\caption{Corpus composition by kind of material. Hours are summed over microphone channels, and overrun is the share of labeled frames with negative torque. Microphone positions apply to the real recordings only. The resyntheses derive from two of the real engines and are not counted as engines.}
\label{tab:corpus}
\centering
\setlength{\tabcolsep}{4pt}
\begin{tabular}{@{}lrrlrr@{}}
\toprule
Material & Hours & Engines & Cylinders & Mic.\ pos. & Overrun \\
\midrule
Real recordings        & 41.9  & 8   & 4, 6, 8 & 14 & 42\,\% \\
Simulated engines      & 60.4  & 370 & 1 to 18 (11) &    & 37\,\% \\
Resyntheses            & 14.4  & (2) & 8            &    & 44\,\% \\
\midrule
Total                  & 116.7 & 378 & 1 to 18 (11) & 14 & 40\,\% \\
\bottomrule
\end{tabular}
\end{table}

\subsection{Recording Conditions and Labels}
\label{subsec:recording}

The real material was recorded on a chassis dynamometer with road-load simulation in a semi-anechoic room. Each session was captured by 2 to 14 microphones at once, all sharing one label track. Across sessions, the microphones occupy 14 distinct positions in the engine bay, at the exhaust and in the cabin.

Labels are recorded with the audio, with no manual annotation: from the vehicle telemetry for the real recordings and the resyntheses, and from the simulation for the simulated engines.

\begin{figure}[t]
\centering
\includegraphics[width=\columnwidth]{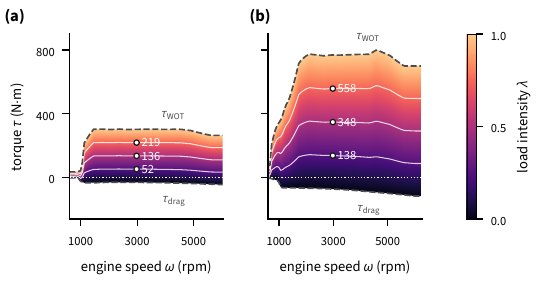}
\input{figures/fig_lambda.caption}
\label{fig:lambda}
\end{figure}

Engine torque is supervised through the load intensity $\load$ (Section~\ref{subsec:signal}):
\begin{equation}
  \load = \clip\!\left(
    \frac{\torque - \torquedrag(\speed)}{\torquewot(\speed) - \torquedrag(\speed)},
    \,0,\,1\right),
  \label{eq:lambda}
\end{equation}
where the full-load ceiling $\torquewot(\speed)$ and the drag floor $\torquedrag(\speed)$ are fitted once per engine on its labeled torque, so that 0 marks the deepest overrun and 1 full load at every speed of every engine (Fig.~\ref{fig:lambda}).

\subsection{Coverage}
\label{subsec:coverage}

\begin{figure}[t]
\centering
\includegraphics[width=\columnwidth]{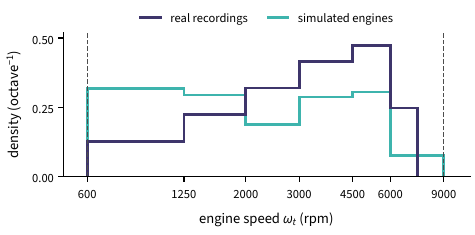}
\input{figures/fig_coverage.caption}
\label{fig:coverage}
\end{figure}

The real and the simulated material vary different factors of engine sound (Section~\ref{subsec:signal}), and so complement each other. The real recordings vary the transfer path: the simultaneous microphones of a session record one labeled state through different paths. The simulated engines vary the engine design, with 11 cylinder counts against three in the real recordings, although five of the eleven counts have only one or two engines each.
Over the operating range, the simulated engines also fill the idle range below $1250\un{rpm}$, where the real recordings hold less than half the frames per octave of the simulated engines (Fig.~\ref{fig:coverage}).
On the dynamometer, the engine can be held at a fixed speed while the load varies. The real material therefore contains steady speed at every load from overrun to full load, the states in which the acceleration reveals nothing about the torque.
One condition is rare in both kinds of material: speed changes faster than $1500\un{rpm/s}$ make up 1.1\,\% of the training frames, and training draws them about three times as often as their share.

\subsection{Split}
\label{subsec:split}

The split is grouped by recording session, so the simultaneous microphones of a session always fall on the same side, and an engine is held out whole where a question below requires it. Resyntheses derived from the test engine are withheld from every split (Table~\ref{tab:split}), and the simulated engines of validation and test are disjoint from those in training. The held-out material answers three questions:
\begin{itemize}
  \item \emph{Does the estimator generalize to an engine type it has never heard?} A real four-cylinder engine of a type absent from training is held out whole in validation. It was also recorded with a setup no training engine shares, so the effects of the two changes cannot be separated.
  \item \emph{Does it transfer to another physical engine of a known type?} The real test material is one eight-cylinder engine, held out whole, whose type enters training only through a second engine of the same type.
  \item \emph{Does it generalize to new recording sessions of an engine it was trained on?} One real eight-cylinder engine contributes some of its recording sessions to validation and the rest to training.
\end{itemize}

\begin{table}[t]
\caption{Material per split, in hours summed over microphone channels. One real engine contributes sessions to both training and validation. Resyntheses of the test engine and simulated renders without a labeled frame (0.3\,h) are in no split.}
\label{tab:split}
\centering
\begin{tabular}{@{}lrrrrr@{}}
\toprule
 & \multicolumn{2}{c}{Real} & \multicolumn{2}{c}{Simulated} & Resyntheses \\
\cmidrule(lr){2-3}\cmidrule(lr){4-5}\cmidrule(l){6-6}
Split & Engines & Hours & Engines & Hours & Hours \\
\midrule
Training   & 6 & 34.8 & 304 & 49.2 & 11.9 \\
Validation & 2 & 2.5  & 26  & 4.3  &      \\
Test       & 1 & 4.6  & 40  & 6.6  &      \\
\bottomrule
\end{tabular}
\end{table}

\section{Experiments}
\label{sec:experiments}

\subsection{Training Details}
\label{subsec:training}

\textbf{Speed scorer:} Every learned weight and every standardization statistic is fitted on the training split, and no statistic is taken per recording.
We minimize $\loss$ with AdamW at a learning rate of $10^{-3}$, 200 warm-up steps and cosine decay to zero over $10\,000$ steps.
Each batch holds 32 chunks of 256 frames ($16.4\un{s}$), and real recordings supply 36.3\,\% of the frames.
The front end and the convolutional stages run in bfloat16 and the scores in fp32.
The output weights $\readout$ of the reader start at zero, so the untrained model returns a uniform posterior.
One run takes $11.5\un{h}$ on one RTX 3090 GPU.

\textbf{Load-intensity head:} The head is fitted after the scorer, over 3000 steps of Adam, with its features taken at the decoded speed, so that it learns under the speed errors it meets at inference.
The absolute level of a recording is set by the microphone, its position and the recording chain, and it carries no load information (Section~\ref{subsec:signal}). During training, the 24 levels of the band profile of a session are therefore shifted together by one random gain of up to $\pm 12\un{dB}$, which leaves the head the level differences between the orders and the ratio of harmonic to noise energy.

\subsection{Evaluation Protocol and Metrics}
\label{subsec:protocol}

The evaluation follows the three questions of Section~\ref{subsec:split}: each defines a slice of the validation and test material, with real and simulated material scored apart. Every metric is computed on held-out recordings, a recording being the signal of one microphone over one session, or one simulated run. Frame metrics are pooled over the labeled frames of a slice, the frames that carry a target (Section~\ref{subsec:objective}).
Rates over recordings, such as the share of broken recordings, count each recording with at least 50 labeled frames once, with equal weight.
All selection uses the validation split, and the test split was evaluated once, after all selection.

\textbf{Engine speed:} The speed metrics measure how much of a recording the estimate tracks closely, how often it falls into the octave confusion of Section~\ref{subsec:signal}, and in which conditions it fails. With the relative error $\relerr_\frameidx = \lvert \speedhat_\frameidx - \speed_\frameidx \rvert / \speed_\frameidx$, we write c5 and c2 for the shares of labeled frames with $\relerr_\frameidx < 0.05$ and $\relerr_\frameidx < 0.02$.
An octave error is a labeled frame whose estimate lies within 0.2 octave of half or of double the true speed.
A wrong count read $\ncylhat$ that keeps the firing frequency scales the estimate by $\ncyl/\ncylhat$. Under such a read, $\ncylhat = 2\ncyl$ or $\ncylhat = \ncyl/2$ produces an octave error, and a wrong count whose ratio lies more than 0.2 octave from a factor of two lowers c5 and c2 without counting as one.
At $3000\un{rpm}$, c2 counts an estimate within $60\un{rpm}$ as correct. We define a recording as \emph{broken} when its c5 lies below 0.2.
Even a correct estimate cannot reach a c2 of 1. The reference speed is sampled every $64\un{ms}$ and jumps within one frame at a gear change or a free rev, while an estimate read through an analysis window passes through such a jump over a frame or two. As a ceiling, the reference smoothed over three frames, $192\un{ms}$, scores c2 0.988 and c5 0.997 on validation, and 0.939 and 0.983 on the held-out sessions, which hold the most such jumps.
Two bands locate failures in the two conditions that strain the spectral resolution (Section~\ref{subsec:signal}). On \emph{ramps}, the frames whose true speed changes by 500 to $1500\un{rpm/s}$ over a centered difference of $\pm 0.5\un{s}$, we report c5. At \emph{idle}, the frames with a true speed below $1250\un{rpm}$, we report c2.
Faster changes come mostly from free revving at standstill and from gear changes, and we report c5 on the frames above $1500\un{rpm/s}$ separately, as the \emph{stress band}.
On the real validation recordings, these frames come almost entirely from one standstill session.
Every speed metric scores the path decoded under the count read from the session (Eq.~\eqref{eq:countread}), with the engine-off gate disabled. Beside it we report the gated variant, in which labeled frames that the gate closes count as misses. The true count serves only as a control, to measure what reading the count costs.

\textbf{Cylinder count:} The count metric tests whether the model names the cylinder count of an engine from the audio of one session, with nothing else known about the engine. It is scored as the share of sessions, or of recordings for the simulated material, whose read count $\ncylhat$ of Eq.~\eqref{eq:countread} equals the cylinder count of the engine. We report the count read over all eleven counts of $\ncylset$ and as the \emph{six-count read}, which takes the arg max of Eq.~\eqref{eq:countread} over the six counts with at least seven training engines, $\ncylsetsix = \{4, 5, 6, 8, 10, 12\}$, on the same scores and paths. Training covers all eleven counts, the held-out engines a subset: the validation engines have 4, 6, 8 and 12 cylinders, and the test engines 2, 4, 5, 6, 8, 10 and 12.

\textbf{Load intensity:} The load metrics describe how well the estimate places the engine between overrun and full load, the position $\load$ of Eq.~\eqref{eq:lambda}. They measure the distance from the true position, the distinction between an engine that drives the vehicle and one driven by it, and the coarse level of load. The distance is the MAE of $\loadhat_\frameidx$ against $\load_\frameidx$, reported with the $R^2$, the share of the variance of $\load_\frameidx$ that the estimate explains. Both are pooled over the frames of a slice on which the engine runs and the speed estimate lies within the grid.
The distinction between driving and overrun is the \emph{state accuracy}: the share of frames on which $\loadstatehat_\frameidx$, thresholded at 0.5, matches the load state $\loadstate_\frameidx$.
For the coarse level, $\load_\frameidx$ and $\loadhat_\frameidx$ are binned into four classes. The overrun class lies below a fixed boundary $\loadcoast = 0.174$, the zero-torque point of an engine whose drag floor is $-0.21\,\torquewot(\speed)$, and three equal classes above it hold light, mid and high load. The fixed boundary keeps the classes the same on every engine, while the state accuracy uses the sign of the torque itself. We report the share of frames in the true class and within one class of it.

\section{Results}
\label{sec:results}

On the two real engines it has never heard, with the microphone position, the engine type and the cylinder count unknown, SonoDyno places 93 to 97\,\% of the frames within 2\,\% of the true engine speed and 94 to 97\,\% within 5\,\%,
identifies the cylinder count with 80\,\% accuracy, read once per recording session,
and confuses the octave on fewer than 2\,\% of the frames.
On the second of them, whose type enters training only through another engine, its load-intensity estimate lies within 0.12 of the true value on average and tells a driving engine from one in overrun on 91\,\% of the frames.
General pitch trackers fall behind even when supplied with the one quantity SonoDyno has to read itself, the harmonic order of every engine: the best of them places 62\,\% of the frames within 5\,\% of the true speed on the real validation recordings, against 96\,\% for SonoDyno (Table~\ref{tab:benchmark}).

\subsection{Engine Speed and Cylinder Count on Validation}
\label{subsec:results_rpm}

Pooled over the validation split, SonoDyno reaches c5 0.977 and c2 0.969, but 388 of the 722 recordings are simulated sweeps, so Table~\ref{tab:speed} reports each held-out slice on its own.

\begin{table}[t]
\caption{Engine speed and cylinder count per held-out slice, on validation and on the test set: c5 and c2 over frames without the engine-off gate and c5 with it (gated), and the count read per session for the real engines and per recording for the simulated ones.}
\label{tab:speed}
\centering
\footnotesize
\setlength{\tabcolsep}{3pt}
\begin{tabular}{@{}lrrrr@{}}
\toprule
Slice & c5 & c2 & gated c5 & count \\
\midrule
\multicolumn{5}{@{}l}{\emph{Validation}} \\
real recordings, pooled & 0.947 & 0.935 & 0.944 &  \\
unseen engine type & 0.969 & 0.968 & 0.969 & 36/47 \\
held-out sessions & 0.894 & 0.859 & 0.887 & 4/4 \\
unseen simulated engines & 0.994 & 0.989 & 0.973 & 60/60 \\
\midrule
\multicolumn{5}{@{}l}{\emph{Test}} \\
unseen engine of a known type & 0.945 & 0.931 & 0.917 & 9/9 \\
unseen simulated engines & 0.980 & 0.974 & 0.966 & 135/149 \\
\bottomrule
\end{tabular}

\end{table}

Accuracy does not follow familiarity. The engine type absent from training, which also carries the change of recording setup, is read more closely (c5 0.969, c2 0.968) than the held-out sessions of a training engine (c5 0.894, c2 0.859), and the unseen simulated engines are read closest (c5 0.994).
The held-out sessions are long dynamic drives, full of the jumps in the reference that a windowed estimate passes through over a frame or two: there, even the smoothed reference scores only c2 0.939 (Section~\ref{subsec:protocol}).

On validation, the cylinder count is misread only on the unseen four-cylinder engine, where 11 of 47 sessions are read as eight cylinders (Table~\ref{tab:speed}). An eight-cylinder hypothesis at the true speed explains the even firing harmonics of a four-cylinder engine, so the decoded speed stays right while the count does not.
Reading the count costs no speed either: decoded under the true count, the same recordings differ by at most 0.003 in c5 and in c2 on every slice. Widening the hypotheses from the six well-covered counts to all eleven changes no session's count, so a rare count never wins.

The remaining errors sit where the analysis needs fine resolution in time and in frequency at once, on ramps and at idle. On the real recordings, each loses about one frame in six to seven (c5 0.840 on ramps, c2 0.832 and c5 0.860 at idle).
The stress band loses one frame in three (c5 0.67), against one in nine on the unseen simulated engines (0.89).
Octave confusion is almost absent. On the real validation recordings of the benchmark, half the true speed takes 3 to 11\,\% of the pitch trackers' frames even under the oracle order, and 0.9\,\% of SonoDyno's; over all labeled validation frames, 0.3\,\% lie at half and 0.5\,\% at double the true speed, 1.8\,\% on the unseen engine type.

\subsection{Benchmark Against Pitch and Periodicity Trackers}
\label{subsec:results_benchmark}

The benchmark runs on a fixed subsample of the validation recordings, 163 of the 334 real and 97 of the 388 simulated ones, on which SonoDyno's values differ from those of Table~\ref{tab:speed}. Table~\ref{tab:benchmark} scores its real recordings, and Table~\ref{tab:benchmark_full} in the appendix every slice, under the oracle-order condition of Section~\ref{subsec:engineaudio}, which hands every tracker the order SonoDyno reads itself.
The nearest tracker, PESTO, places 61.6\,\% of the frames within 5\,\% against 95.7\,\% for SonoDyno, and 56.9\,\% within 2\,\% against 94.8\,\%.
Its errors are gross, a mean relative error of 64\,\% against 2.4\,\%, the signature of locking onto the wrong harmonic. CREPE, pYIN and YIN stay below 40\,\% of the frames.
The order representation alone outperforms every tracker. With every learned stage removed, SonoDyno's front end and decoder alone, the order energy of the three windows averaged and summed flat over the 24 orders, place 72\,\% of the frames within 5\,\% and 71\,\% within 2\,\%. Weighted towards the firing orders of the true count, they reach 77\,\% and 76\,\%. Their errors are mostly octave errors below $3000\un{rpm}$, and the learned stages add the rest, up to 95.7\,\% of the frames within 5\,\% and 94.8\,\% within 2\,\%, with broken recordings down from 30\,\% to 0.6\,\% (Table~\ref{tab:benchmark}).

\begin{table}[t]
\caption{Engine speed on the 163 real recordings of the benchmark subsample of validation against general pitch and periodicity trackers, every tracker under the oracle order, SonoDyno and its variant trained on simulated data only with the count read from the audio. c5, c2 and RMAE, the mean relative error, are taken over frames, broken over recordings. The retrained PESTO rows cover 153 of the 163 recordings. The best value of each column is in bold.}
\label{tab:benchmark}
\centering
\footnotesize
\setlength{\tabcolsep}{3pt}
\begin{tabular}{@{}lrrrr@{}}
\toprule
System & c5 & c2 & RMAE \% & broken \% \\
\midrule
SonoDyno & \textbf{0.957} & \textbf{0.948} & \textbf{2.4} & \textbf{0.6} \\
SonoDyno, simulated data only & 0.638 & 0.616 & 19.2 & 33.1 \\
no learned stage, orders flat & 0.718 & 0.709 & 19.9 & 30.1 \\
no learned stage, true firing orders & 0.774 & 0.764 & 20.0 & 20.2 \\
PESTO & 0.616 & 0.569 & 63.8 & 17.2 \\
CREPE & 0.378 & 0.342 & 206.4 & 33.7 \\
pYIN & 0.258 & 0.253 & 78.8 & 62.6 \\
YIN & 0.279 & 0.275 & 81.1 & 55.2 \\
PESTO retrained, our audio & 0.075 & 0.031 & 52.0 & 88.9 \\
PESTO retrained, eight cylinders & 0.115 & 0.034 & 60.3 & 90.2 \\
\bottomrule
\end{tabular}

\end{table}

Retrained self-supervised on 23\,h of our engine audio, as Motor2Synth does on its own data, PESTO becomes worse than the stock model and places 8\,\% of the frames within 5\,\% under the oracle order (Table~\ref{tab:benchmark}). Restricting the retraining to the real eight-cylinder engines recovers nothing (12\,\%), and neither does the octave correction of the published recipe. Without the anchor of Section~\ref{subsec:pitch}, the model locks onto no order at all.

Read as the rotation frequency, with the speed taken as sixty times the tracked pitch, every tracker falls below 14\,\% of the frames. Each locks onto a harmonic whose order changes with the engine and within a recording. Choosing the order per engine lifts PESTO from 6 to 62\,\%.

The speed scorer is no larger than the smallest learned tracker: it has 124\,302 learned parameters against 130\,806 for PESTO and 22.2\,M for CREPE, and the load head adds 164\,649.

\subsection{Load Intensity}
\label{subsec:results_lam}

The load intensity is readable from the sound. On the held-out sessions of the training engine the estimate explains 0.76 of its variance with a mean absolute error of 0.09, a third of the width of one coarse class of Section~\ref{subsec:protocol} (Table~\ref{tab:load}). It reads the sign of the torque, driving against overrun, on 91\,\% of the frames, more reliably than its level, with the right coarse class on 76\,\%.
The microphones of a session agree: their estimates correlate at 0.93, and averaging them (Section~\ref{subsec:loadhead}) raises the explained variance from 0.76 to 0.82.

\begin{table}[t]
\caption{Load intensity on validation and on the test set: explained variance and mean absolute error on the axis from overrun to full load, state accuracy and the share of frames in the true coarse class, per microphone and averaged over the microphones of a session.}
\label{tab:load}
\centering
\footnotesize
\begin{tabular}{@{}lrrrr@{}}
\toprule
Slice & $R^2$ & MAE & state & class \\
\midrule
\multicolumn{5}{@{}l}{\emph{Validation}} \\
held-out sessions, training engine & 0.756 & 0.090 & 0.905 & 0.761 \\
\quad averaged over microphones & 0.819 & 0.082 & 0.929 & 0.795 \\
unseen simulated engines & 0.819 & 0.108 & & 0.779 \\
\midrule
\multicolumn{5}{@{}l}{\emph{Test}} \\
unseen engine of a known type & 0.640 & 0.117 & 0.914 & 0.697 \\
\quad averaged over microphones & 0.781 & 0.099 & 0.954 & 0.718 \\
unseen simulated engines & 0.785 & 0.114 & 0.938 & 0.780 \\
\bottomrule
\end{tabular}

\end{table}

On the unseen simulated engines, 26 engines of several counts and capacities (Table~\ref{tab:split}), the estimate explains 0.82 of the variance with an error of 0.11: the target normalized to each engine's own capacity carries the reading to engines the head has never heard.

Training the same head at the labeled speed in place of the decoded one drops the explained variance on the held-out sessions from 0.76 to 0.48: the head has to learn under the speed errors it meets at inference (Section~\ref{subsec:training}).

\subsection{Test Set}
\label{subsec:results_test}

The test engine answers the second question of Section~\ref{subsec:split}. It is read at the level of the engine type absent from training (c5 0.945 against 0.969) and above the held-out sessions. Its count is right in all nine sessions, and decoding under the true count gives the same c5 (Table~\ref{tab:speed}).
The bands repeat the validation picture: ordinary accelerations keep c5 0.883, idle c5 0.858 against 0.860 on validation, and the stress band drops to 0.747.
The gap to the unseen engine type follows from the sessions: long drives that spend more of their labeled frames than the real training material at idle, where the speed is hardest to read (21 against 12\,\%), and in overrun, where the orders sink towards the noise floor (54 against 41\,\%).

For the load, the test microphones disagree more than on validation, and averaging over them recovers more. Per microphone the estimate explains 0.64 of the variance, and the session average 0.78, within 0.04 of the validation engine's 0.82. The error falls from 0.12 to 0.10, and the state accuracy rises from 0.91 to 0.95 (Table~\ref{tab:load}).
The misses are to a neighboring class, 70\,\% exact and 93\,\% within one class, and frames in overrun read at 0.08, a margin of 0.09 below the class boundary of 0.174.

The unseen simulated test engines reproduce their validation counterparts within 0.015 of c5 (0.980 against 0.994), 0.015 of c2 (0.974 against 0.989) and 0.04 of explained variance (0.78 against 0.82), on a set that adds the counts 2, 5 and 10. The count is right on 135 of 149 simulated recordings (Tables~\ref{tab:speed} and~\ref{tab:load}).

\subsection{Design Choices and Negative Results}
\label{subsec:design_choices}

The variants below are trained alike and decoded under one fixed count per engine, so they differ only in the scorer (Table~\ref{tab:ladder}). The variants of the training data, the schedule and the decoder use the count read from the audio, as in Section~\ref{subsec:results_rpm}.
Two seeds of one architecture differ by up to 0.06 on the held-out sessions' stress band, and smaller single-seed differences are not interpreted.

\textbf{Scorer components:} The order-pattern detector carries the dynamics, and the count-weighted orders alone carry steady speed. Zeroing the detector in the trained model drops c5 in the held-out sessions' stress band from 0.51 to 0.10 and c2 at idle from 0.81 to 0.52, while the recordings of the unseen engine type keep c5 0.91 of their 0.96.
With the same weighted evidence, the MLP reader beats a plain weighted sum of the orders: it lifts the count read on the unseen engine type from 21 to 37 of 47 sessions and the stress band from 0.22 to 0.36.
The reader and the detector together take c2 over all validation from 0.937 to 0.957.
The orders have to stay apart until the firing-order gain weights them. A variant that mixes them early, with a shared trunk, anonymous channels per order, the gain applied to those channels and a plain sum over the windows, plateaus at c5 0.926, with the stress band at 0.11 and a count read of 20 of 47 sessions. The proposed architecture reaches 0.965, 0.51 and 37 of 47 at the same step count.

\textbf{Capacity:} A refinement twice as wide and one block deeper does not exceed the proposed sizing (c5 0.973 against 0.976), at higher training cost.
Removing channels from that wide model at inference locates its capability. Four of its eight refinement channels per window carry it, while keeping only 8 of the 32 fusion channels drops the stress band from 0.57 to 0.12 on the held-out sessions and from 0.86 to 0.16 on the unseen simulated engines: the fusion width carries the stress band.

\textbf{Training data and schedule:} The oversampling of fast frames of Section~\ref{subsec:coverage} lifts the held-out sessions' stress band from 0.60 to 0.68 and their c5 from 0.87 to 0.89, and the ordinary accelerations of all real recordings from 0.81 to 0.84. It moves the unseen engine type by less than the seed spread.
Raising the share of real recordings among the training frames from 36.3 to 53\,\%, by contrast, fits the known engines and costs the unknown one: the held-out sessions' c5 rises by 0.010, while the unseen engine type loses 0.023 of c5 and five of its 36 correct count reads.
Trained on simulated engines alone, the same recipe matches SonoDyno on unseen simulated engines (c5 0.995 against 0.994) but reaches 0.638 on the real recordings, below the front end without a learned stage, and 0.822 given the true count (Table~\ref{tab:benchmark}, Table~\ref{tab:benchmark_full}): the real recordings carry the count read on real engines and part of the speed read. Its load-intensity head, fitted on simulated engines alone, does not carry over to real engines either.
Twice the training steps do not help: the unseen engine type loses 0.010 of c5, and the share of its idle frames estimated at double the speed rises from 5.0 to 7.4\,\%.

\textbf{Further variants:} Allowed to go negative, the weights of the firing-order gain stay almost non-negative, with a floor of about $-0.035$, and the model stays within the seed spread of its baseline in c5 over all validation. The gain therefore weights evidence that an order is present.
An adaptive transition cost in the manner of ARC~\cite{huaARC2026}, which scales the decoder's acceleration cost by the posterior confidence of the frame, lowers c5 by 0.0015 on the real recordings and 0.0014 on the unseen engine type, and the real stress band by 0.006. The decoder keeps a fixed cost.
Four evidence channels per order in place of one, and weight decay, both lower the count read and c2 on the real recordings.

\section{Discussion}
\label{sec:discussion}

\subsection{Relation to Order Tracking}
Tacholess order tracking produces a speed trace for a known machine. The harmonic order is chosen by hand or per data set, the trajectory is assumed continuous from an initial estimate, and the sensor is an accelerometer on the structure~\cite{luTacholessSpeedEstimation2019, peetersReviewComparisonTacholess2019}. A review of the field names low speeds, changes of regime and fast fluctuations as its open challenges~\cite{peetersReviewComparisonTacholess2019}. SonoDyno starts cold on an unknown engine, from an airborne microphone: it identifies the order through the count read and decodes the speed under the rotational dynamics of the engine. Of the three challenges, it meets the low speeds in part, with c2 0.83 at idle on the real recordings, and reads across the change of harmonic structure between load and overrun, because both states are in its training material. The fastest fluctuations remain open: the stress band is read at two thirds of its frames (Section~\ref{subsec:results_rpm}).
The contemporaneous tracker ARC also places its contribution in the decoder and keeps the octave ambiguity for a downstream decision~\cite{huaARC2026}, an ambiguity SonoDyno resolves in the joint score. An adaptive transition cost in the manner of ARC, moved into our decoder, gains only on simulated sweeps (Section~\ref{subsec:design_choices}).

\subsection{Relation to Pitch Estimation}
Harmonic-aware pitch estimators build in what the front end makes explicit. The harmonic constant-Q transform stacks harmonics as channels on one logarithmic axis~\cite{bittnerDeepSalience2017}, and harmonic dilations place a kernel's taps on them~\cite{weiHarmoF02022}. The order comb does the same for the engine orders. Those models fix the harmonic index to the integers of a musical fundamental, while SonoDyno treats it as a hypothesis, read through the firing-order gain. Two further components have no counterpart in pitch trackers. One refines each harmonic within each resolution and fuses the resolutions per harmonic before any read across harmonics. The other is a decoder whose state carries the rate of change of the path. Both are small and could serve beyond engine audio, but we have not tested them on music or speech.

The need for an anchor does carry over. A relative objective learns pitch up to a shift~\cite{gfellerSPICE2020, riouPESTO2023}. Engine audio has no fixed anchor, because the strongest line is an order that changes with the engine. The count read supplies it. The retrained PESTO of Section~\ref{subsec:results_benchmark} shows what happens without it. Cross-domain pitch estimation holds where the training range covers the test range~\cite{morrisonCrossDomainNeuralPitch2023}, and the engine's fundamentals lie below the ranges those models were built for.

\subsection{Load Without Telemetry}
The synthesis and sound-design systems of Section~\ref{subsec:engineaudio} read the load from the bus or a dynamometer, or substitute loudness. Bus access is often proprietary, older vehicles have no bus, and a dynamometer is no option for a recording that already exists. The recordings that exist in volume, in sample packs, archives, field and video recordings, carry no state at all. From such a recording alone, SonoDyno reads an engine-normalized effort. Its sign is the most reliable part. Training under a random level shift leaves the head the level differences between the orders and the harmonic-to-noise ratio (Section~\ref{subsec:training}). Part of the error differs from microphone to microphone, and averaging over the microphones of a session removes it (Section~\ref{subsec:results_test}). For noise, vibration and harshness work and for machinery analysis, this accuracy and the absence of an absolute torque fall short. For conditioned synthesis, however, the effort is the control the models need, more general than loudness, and it turns that volume of unannotated audio into training material.

\subsection{Limitations}
The remaining speed errors sit on ramps, in the stress band and at idle.

Real engines contribute only four, six and eight cylinders, and the held-out real engines only four and eight, so every other count, six included, is tested on simulated engines alone. The one systematic miss, a four-cylinder engine at constant high speed read as eight, is a confusion the signal model allows.

On the unseen engine type, the change of engine and the change of setup are not separated (Section~\ref{subsec:split}). All real material was recorded on chassis dynamometers in semi-anechoic rooms, so road, wind and traffic noise are not covered.

The speed estimate rests on one training seed, with the seed spread given in Section~\ref{subsec:design_choices}. The test split was evaluated once, on one real engine. The load estimate is scored on held-out sessions of a training engine, on simulated engines and on the test engine, and on no real engine of an unseen type.

\section{Conclusion}
\label{sec:conclusion}

SonoDyno estimates the operating state of a combustion engine from a microphone signal, with no tachometer, no vehicle bus and no prior knowledge of the engine. It reads the cylinder count together with the speed, and reading the count costs the speed nothing. It estimates the load as an effort normalized to the engine's own capacity.

Both estimates rest on physics built into the representation. On a logarithmic speed grid, the order comb turns the harmonic structure of engine sound into order energy, on which the decoder alone, with no learned stage, already outperforms every general pitch tracker. The count enters only as weights over the orders, so one model serves every count. The decoder charges changes of the engine's acceleration, which follows the slowly changing net torque. The load target places the torque within the engine's own range, from its drag floor to its full-load ceiling, a scale the sound carries and loudness does not.

A recording without telemetry can therefore be annotated with the speed and load that conditioned synthesis and sound design need, which turns the engine recordings available in volume into training material. Beyond engines, whether the per-harmonic refinement, the fusion of resolutions and the rotational-dynamics decoder serve pitch estimation remains open.

\appendix[Benchmark per Slice and Ablation Variants]
\label{app:benchmark}
Table~\ref{tab:benchmark_full} repeats the benchmark of Table~\ref{tab:benchmark} on every validation slice, under the same conditions. Table~\ref{tab:ladder} lists the ablation variants of Section~\ref{subsec:design_choices}.

\begin{table}[t]
\caption{The benchmark of Table~\ref{tab:benchmark} on the subsample of every validation slice, with the number of recordings in parentheses. c5, c2 and RMAE, the mean relative error, are taken over frames, broken over recordings.}
\label{tab:benchmark_full}
\centering
\footnotesize
\setlength{\tabcolsep}{3pt}
\begin{tabular}{@{}lrrrr@{}}
\toprule
System & c5 & c2 & RMAE \% & broken \% \\
\midrule
\multicolumn{5}{@{}l}{\emph{Real and unseen simulated (260)}} \\
SonoDyno & \textbf{0.974} & \textbf{0.967} & \textbf{1.7} & \textbf{0.4} \\
SonoDyno, simulated data only & 0.805 & 0.791 & 10.4 & 20.8 \\
no learned stage, orders flat & 0.622 & 0.613 & 66.1 & 25.0 \\
no learned stage, true firing orders & 0.637 & 0.628 & 89.4 & 18.1 \\
PESTO & 0.546 & 0.508 & 87.6 & 14.6 \\
CREPE & 0.484 & 0.454 & 147.5 & 23.1 \\
pYIN & 0.405 & 0.384 & 70.1 & 44.2 \\
YIN & 0.408 & 0.391 & 81.0 & 40.4 \\
PESTO retrained, our audio & 0.101 & 0.041 & 43.8 & 89.6 \\
PESTO retrained, eight cylinders & 0.102 & 0.035 & 54.0 & 93.2 \\
\midrule
\multicolumn{5}{@{}l}{\emph{Real recordings (163)}} \\
SonoDyno & \textbf{0.957} & \textbf{0.948} & \textbf{2.4} & \textbf{0.6} \\
SonoDyno, simulated data only & 0.638 & 0.616 & 19.2 & 33.1 \\
no learned stage, orders flat & 0.718 & 0.709 & 19.9 & 30.1 \\
no learned stage, true firing orders & 0.774 & 0.764 & 20.0 & 20.2 \\
PESTO & 0.616 & 0.569 & 63.8 & 17.2 \\
CREPE & 0.378 & 0.342 & 206.4 & 33.7 \\
pYIN & 0.258 & 0.253 & 78.8 & 62.6 \\
YIN & 0.279 & 0.275 & 81.1 & 55.2 \\
PESTO retrained, our audio & 0.075 & 0.031 & 52.0 & 88.9 \\
PESTO retrained, eight cylinders & 0.115 & 0.034 & 60.3 & 90.2 \\
\midrule
\multicolumn{5}{@{}l}{\emph{Unseen engine type, real (141)}} \\
SonoDyno & \textbf{0.961} & \textbf{0.960} & \textbf{2.3} & \textbf{0.7} \\
SonoDyno, simulated data only & 0.615 & 0.599 & 19.1 & 35.5 \\
no learned stage, orders flat & 0.700 & 0.698 & 22.1 & 31.2 \\
no learned stage, true firing orders & 0.768 & 0.767 & 22.4 & 19.9 \\
PESTO & 0.717 & 0.659 & 56.1 & 13.5 \\
CREPE & 0.388 & 0.344 & 132.0 & 32.6 \\
pYIN & 0.204 & 0.203 & 78.9 & 66.7 \\
YIN & 0.230 & 0.228 & 74.9 & 58.2 \\
PESTO retrained, our audio & 0.075 & 0.030 & 53.1 & 88.6 \\
PESTO retrained, eight cylinders & 0.102 & 0.033 & 62.5 & 91.5 \\
\midrule
\multicolumn{5}{@{}l}{\emph{Held-out sessions of a training engine (22)}} \\
SonoDyno & \textbf{0.948} & \textbf{0.920} & \textbf{2.5} & \textbf{0.0} \\
SonoDyno, simulated data only & 0.690 & 0.658 & 19.3 & 18.2 \\
no learned stage, orders flat & 0.757 & 0.733 & 14.9 & 22.7 \\
no learned stage, true firing orders & 0.789 & 0.758 & 14.5 & 22.7 \\
PESTO & 0.384 & 0.360 & 81.4 & 40.9 \\
CREPE & 0.354 & 0.337 & 378.0 & 40.9 \\
pYIN & 0.381 & 0.369 & 78.4 & 36.4 \\
YIN & 0.393 & 0.381 & 95.3 & 36.4 \\
PESTO retrained, our audio & 0.074 & 0.034 & 46.3 & 91.7 \\
PESTO retrained, eight cylinders & 0.181 & 0.042 & 49.0 & 75.0 \\
\midrule
\multicolumn{5}{@{}l}{\emph{Unseen simulated engines (97)}} \\
SonoDyno & 0.994 & 0.989 & 0.8 & \textbf{0.0} \\
SonoDyno, simulated data only & \textbf{0.995} & \textbf{0.990} & \textbf{0.5} & \textbf{0.0} \\
no learned stage, orders flat & 0.513 & 0.504 & 118.7 & 16.5 \\
no learned stage, true firing orders & 0.479 & 0.472 & 168.5 & 14.4 \\
PESTO & 0.466 & 0.439 & 114.9 & 10.3 \\
CREPE & 0.604 & 0.583 & 80.3 & 5.1 \\
pYIN & 0.573 & 0.534 & 60.2 & 13.4 \\
YIN & 0.554 & 0.524 & 81.0 & 15.5 \\
PESTO retrained, our audio & 0.125 & 0.050 & 36.0 & 90.7 \\
PESTO retrained, eight cylinders & 0.089 & 0.037 & 47.9 & 97.9 \\
\bottomrule
\end{tabular}

\end{table}

\begin{table}[t]
\caption{Ablation variants of Section~\ref{subsec:design_choices}, trained for 4000 steps unless stated, each at its final training step, and decoded under one fixed count per engine. Columns: c5 and c2 over all validation, c2 at idle on the real recordings, c5 in the stress band on the held-out sessions of the training engine, and the count read on the unseen engine type as correct sessions out of 47. The last row is the proposed model of Tables~\ref{tab:speed} to~\ref{tab:load}; the other rows are separately trained variants.}
\label{tab:ladder}
\centering
\footnotesize
\setlength{\tabcolsep}{3pt}
\begin{tabular}{@{}lrrrrr@{}}
\toprule
Variant & c5 & c2 & idle c2 & stress & count \\
\midrule
early order mixing, shared trunk & 0.926 & 0.917 & 0.764 & 0.114 & 20/47 \\
per-order chain, sum reader & 0.946 & 0.937 & 0.773 & 0.216 & 21/47 \\
\quad + MLP reader & 0.948 & 0.938 & 0.790 & 0.355 & 37/47 \\
\quad + detector & 0.963 & 0.954 & 0.789 & 0.474 & 28/47 \\
\quad + both & 0.965 & 0.957 & 0.811 & 0.506 & 37/47 \\
\quad both, detector zeroed & 0.752 & 0.745 & 0.516 & 0.105 & -- \\
\quad both, seed 2 & 0.967 & 0.959 & 0.816 & 0.473 & 34/47 \\
signed gain weights & 0.962 & 0.952 & 0.818 & 0.576 & 37/47 \\
wide refinement, 10k steps & 0.973 & 0.965 & 0.820 & 0.603 & 33/47 \\
proposed sizing, 10k steps & 0.976 & 0.969 & 0.828 & 0.597 & 35/47 \\
\quad + fast-frame oversampling & 0.977 & 0.969 & 0.832 & 0.675 & 36/47 \\
\bottomrule
\end{tabular}

\end{table}

\section*{Code and Data Availability}
Upon acceptance, code, the weights of the model trained on simulated data alone (Section~\ref{subsec:design_choices}) and the simulated part of the training data will be released. The real recordings are proprietary and are not released, nor is the model trained on them.

\bibliographystyle{IEEEtran}
\bibliography{refs}

\end{document}